\documentclass[aps,prd,10pt,tightenlines,nofootinbib,superscriptaddress]{revtex4}

\usepackage{amsfonts,amssymb,amsthm,bbm,amsmath}
\usepackage{color,psfrag}
\usepackage[dvips]{graphicx}

\newcommand{\C}{{\mathbb C}}
\newcommand{\N}{{\mathbb N}}
\newcommand{\R}{{\mathbb R}}

\newcommand{\cE}{{\mathcal E}}

\newcommand{\cJ}{{\mathcal J}}

\newcommand{\cV}{{\mathcal V}}
\newcommand{\cR}{{\mathcal R}}
\newcommand{\cH}{{\mathcal H}}

\newcommand{\cT}{{\mathcal T}}
\newcommand{\cD}{{\mathcal D}}

\newcommand{\cO}{{\mathcal O}}
\newcommand{\cW}{{\mathcal W}}
\newcommand{\cS}{{\mathcal S}}
\newcommand{\SU}{\mathrm{SU}}

\newcommand{\SL}{\mathrm{SL}}

\newcommand{\U}{\mathrm{U}}

\newcommand{\be}{\begin{equation}}
\newcommand{\ee}{\end{equation}}
\newcommand{\beq}{\begin{eqnarray}}
\newcommand{\eeq}{\end{eqnarray}}
\newcommand{\bea}{\begin{eqnarray}}
\newcommand{\eea}{\end{eqnarray}}
\newcommand{\nn}{\nonumber}

\newcommand{\mat} [2] {\left ( \begin{array}{#1}#2\end{array} \right ) }

\newcommand{\su}{{\mathfrak su}}

\renewcommand{\u}{{\mathfrak u}}

\newcommand{\la}{\langle}
\newcommand{\ra}{\rangle}

\DeclareMathOperator{\tr}{Tr}
\newcommand{\f}{\frac}
\newcommand{\tl}{\widetilde}
\def\nn{\nonumber}
\def\pp{\partial}
\def\arr{\rightarrow}

\def\vphi{\varphi}
\def\eps{\epsilon}

\newcommand{\id}{\mathbb{I}}

\def\bz{\bar{z}}
\def\bZ{\bar{Z}}
\def\tF{\widetilde{F}}
\def\tG{\widetilde{G}}
\def\bF{\overline{F}}
\def\bE{\overline{E}}
\def\vV{\vec{V}}
\def\vJ{\vec{J}}
\def\vsigma{\vec{\sigma}}
\def\hE{\hat{E}}
\def\hF{\hat{F}}
\def\hFd{{\hat{F}^\dag}}
\def\tT{\widetilde{T}}

\newtheorem{theorem}{Theorem}[]

\newtheorem{res}[theorem]{Result}

\begin{document}

\title{Generating Functions for Coherent Intertwiners}

\author{{\bf Valentin Bonzom}}\email{vbonzom@perimeterinstitute.ca}
\affiliation{Perimeter Institute, 31 Caroline St N, Waterloo ON, Canada N2L 2Y5}
\author{{\bf Etera R. Livine}}\email{etera.livine@ens-lyon.fr}
\affiliation{Laboratoire de Physique, ENS Lyon, CNRS-UMR 5672, 46 All\'ee d'Italie, Lyon 69007, France}
\affiliation{Perimeter Institute, 31 Caroline St N, Waterloo ON, Canada N2L 2Y5}

\date{\today}

\begin{abstract}
We study generating functions for the scalar products of SU(2) coherent intertwiners, which can be interpreted as coherent spin network evaluations on a 2-vertex graph. We show that these generating functions are exactly summable for different choices of combinatorial weights. Moreover, we identify one choice of weight distinguished thanks to its geometric interpretation. As an example of dynamics, we consider the simple case of SU(2) flatness and describe the corresponding Hamiltonian constraint whose quantization on coherent intertwiners leads to partial differential equations that we solve. Furthermore, we generalize explicitly these Wheeler-DeWitt equations for SU(2) flatness on coherent spin networks for arbitrary graphs.

\end{abstract}

\maketitle
\tableofcontents

\section*{Introduction}

Loop quantum gravity is an approach to quantizing general relativity where excitations are carried by embedded graphs so that the kinematical Hilbert space is spanned by (diffeomorphism equivalence classes of) such graphs. When restricted to a single graph, the kinematics is equivalent to lattice $\SU(2)$ gauge theory and thus can be derived from a phase space with Wilson lines on links and their conjugate $\su(2)$-valued electric fluxes. Moreover this phase space admits a natural interpretation in terms of discrete geometries known as twisted geometries \cite{twisted1, polyhedron}.

This is the frame where the dynamics has to be formulated and therefore the quantization heavily relies on $\SU(2)$ representation theory. Typical objects from $\SU(2)$ re-coupling theory are the Wigner 3nj-symbols, which depend on $3n$ angular momenta (spins) built from sums of products of Clebsch-Gordan coefficients \cite{varshalovich}. They arise in (loop) quantum gravity as evaluations of the spin network states of geometry on the trivial connection and as the building blocks of the spinfoam transition amplitudes between those states.

While some basic properties have been known for several decades, a need for new results involving more and more spins have appeared and have led to some interesting progress. They come from various areas of physics, like quantum information \cite{spinnets-marzuoli, 3nj-marzuoli}, semi-classical approximations for quantum angular momenta \cite{littlejohn, Yu}, and quantum gravity \cite{6jnlo, pushing6j, 6jmaite, barrett-asym-summary, recursion, recursion6j_bis, 3njsmall, dowdall-handlebodies}.

While these spin network evaluations are very complicated, it has been noticed that they admit generating functions which are remarkably simple and can be written in a closed form. Schwinger calculated in his seminal paper \cite{schwinger:52} some generating function for the Wigner 6j and 9j-symbols. Bargmann then derived them again through a different reasoning in \cite{bargmann:62} using Gaussian integrals. Since then, generating functions for generic symbols have been evaluated, mostly on algebraic grounds \cite{wu-9j, huang-wu-2j-15j, labarthe, schnetz, garoufalidis, costantino-generating}.

It has been recently understood that such generating functions are very useful for quantum gravity \cite{recursion_spinor, jeff}. Indeed the discrete geometry of loop gravity states - twisted geometries - can be formulated classically with spinors \cite{twisted2}, which are quantized as Schwinger's bosonic operators. This way, loop quantum gravity wave-functions can be represented in a basis of coherent states \cite{spinor, spinor_johannes, un0, un1, un2, un3, un4, un4_conf}. In \cite{recursion_spinor}, it was shown that the wave-function for a flat geometry on the boundary of a tetrahedron (in the context of three-dimensional gravity) is just the Schwinger's generating function for 6j-symbols when written in the appropriate coherent basis.

This result is exciting for the future. Indeed it means that working with some coherent state basis, one trades spin network evaluations to their generating functions, which are holomorphic functions of classical spinors. As often, we expect generating functions to be easier to handle than the symbols themselves. The usual difficulties can be translated into problems of complex analysis. For instance the asymptotic regime of Wigner symbols is hidden in the poles of the Schwinger's generating functions.

Moreover, simple quantum gravity dynamics and aspects of more realistic dynamics have been formulated in terms of recursion equations on the amplitudes \cite{recursion}. For example, 3d gravity and the topological 4d BF model admit Wheeler-DeWitt equations which are difference equations solved by Wigner symbols \cite{recursion3d, recursion4d}. When re-expressing those equations in a spinor coherent basis, they become partial differential equations \cite{recursion_spinor}.

However, those partial differential equations may be quite complicated. They actually depend on a choice of basis of coherent states, corresponding to a choice of combinatorial weights in the definition of the generating functions. The Schwinger's choice which has been used so far in the literature is certainly a good choice to re-sum spin network evaluations for generic graphs as shown in \cite{jeff}, but some other choices may lead to simpler partial differential equations and saddle points with more straightforward geometric interpretation.

We investigate those ideas in the present paper using the special case of the graph with two vertices connected by $N$ links. It is obviously a good testing ground, already considered in \cite{2vertex}, but it is also interesting in itself because the generating functions in this case generate scalar products of $N$-valent intertwiners, which are central objects in quantum gravity.

In Sec. \ref{sec:coh-intertwiners} we review the spinorial description of the LQG phase space and its quantization, presenting different bases of coherent intertwiners. In Sec. \ref{sec:GF-2vertex} we start to focus on the 2-vertex graph and show that the spin network evaluation in such coherent states basis is a generating function for the intertwiner scalar products, and can be written in terms of $\SU(2)$-invariant variables (cross-ratios). We also introduce several choices of combinatorial weights for the generating function, corresponding to different choices of coherent intertwiners.


In Sec. \ref{sec:GF-calculations} we show that it is actually possible to calculate \emph{exactly} these generating functions for \emph{different} choices of combinatorial weights, at least in the case of the 2-vertex graph. We discuss the geometric content of their saddle point evaluations in Sec. \ref{sec:saddlepoint}, where it turns out that a natural geometric interpretation comes out when using a specific choice of weight which we will call the \emph{geometric} choice.

The $\SU(2)$-flat dynamics (as in 3d gravity) is considered in Sec. \ref{sec:pde} and it is found that the simplest Wheeler-DeWitt equation is obtained when considering the geometric generating function. Finally we give some preliminary calculations of generating functions for arbitrary graphs, and in particular obtain derive the Wheeler-DeWitt equations of flat dynamics.

The appendix \ref{gaussian} contains material on constrained Gaussian integrals, and \ref{sec:GF-wigner3nj} relates generating functions of scalar products of coherent intertwiners to generating functions of Wigner symbols.





\section{Coherent Intertwiners and $\U(N)$ Structure} \label{sec:coh-intertwiners}

We present a quick review of the spinorial framework for $\SU(2)$ intertwiners as developed in \cite{un2,un3,un4,spinor}, following the previous identification of an action of the unitary group $\U(N)$ on the space of intertwiners \cite{un0,un1}. In this setting, intertwiners appear as the quantization of classical polyhedra. We start by reviewing the spinor variables for polyhedra and their classical phase space. We then review their quantization into intertwiner states and the operators acting on the intertwiner space. This leads us to the definition of coherent intertwiners. 

\subsection{Spinors and Classical Setting for Intertwiners}

In the following, we call a spinor a complex 2-vector $|z\ra\in\C^2$, living in the fundamental representation of $\SU(2)$, and we define\footnote{Later we also use the matrix $\epsilon = \left(\begin{smallmatrix} 0&1\\-1 &0\end{smallmatrix}\right) = -\varsigma$. Notice also $\varsigma^{-1}=-\varsigma$.} its dual spinor $|z]$
\be
|z\ra=\mat{c}{z^0 \\z^1},\qquad
|z]\equiv\varsigma\,|z\ra
=\mat{cc}{0& -1 \\ 1 &0}\mat{c}{\bz^0 \\ \bz^1}
=\begin{pmatrix} -\bz^1 \\ \bz^0\end{pmatrix}\,.
\ee
We provide $\C^2$ with the natural symplectic structure defined by the canonical Poisson bracket:
\be
\{z_A,\bz_B\}=-i\delta_{AB}\,.
\ee
We further define the 3-vector $\vV(z)\in\R^3$ obtained by projecting the spinor on the Pauli matrices:
\be
\vV(z)=\la z|\vsigma|z\ra,\qquad
|z\ra\la z|=\f12\left(\la z|z\ra+\vV(z)\cdot\vsigma\right),
\ee
where the Pauli matrices are normalized such that $\tr\,\sigma_a\sigma_b=\,2\delta_{ab}$ for $a$ and $b$ running from 1 to 3. Whenever there is no confusion, we will omit the argument $z$ for $\vV$. This vector has norm $|\vV|=\la z|z\ra$ and completely determines the spinor $z$ up a global phase. Moreover, the Poisson brackets of its components define a $\su(2)$ algebra:
\be
\{V_a,V_b\}=2\eps_{abc}V_c.
\ee
Actually $\vV$ generates the action of $\SU(2)$ on the spinors, $|z\ra\arr g\,|z\ra$ for $g\in\SU(2)$.

Now, the phase space for intertwiners with $N$ legs is defined by $N$ spinors $z_i$ living in $\C^{2N}$ satisfying the closure constraints:
\be
\sum_i |z_i\ra\la z_i|=\f12\sum_i \la z_i|z_i\ra\,\id
\quad\textrm{or equivalently}\quad
\sum_i \vV_i =0\,.
\ee
These are first class constraints generating global $\SU(2)$ transformations on all $N$ spinors. From a geometrical perspective, the constraint $\sum_i \vV_i=0$ implies that the $N$ vectors $\vV_i$ define a unique convex polyhedron with $N$ faces: the vectors $\vV_i$ are the normal vectors to the faces. For more details on the reconstruction of this dual polyhedron, the interested reader is referred to \cite{polyhedron}.

Thus our intertwiner phase space is defined by the symplectic reduction by the closure constraints: $\C^{2N}//\SU(2)$ is the space of spinors satisfying the closure constraints and up to global $\SU(2)$ transformations. It describes the set of framed polyhedra with $N$ faces \cite{un1,un2}. By ``framed", we mean that we have an extra $\U(1)$ phase attached to each face, which is the degree of freedom contained in $z_i$ compared to $\vV_i$. These faces are mostly irrelevant when studying single intertwiners, but are needed when gluing those intertwiners into spin network states.

\smallskip

Before moving to the quantization and to intertwiner states, let us further define $\SU(2)$-invariant observables on the constrained phase space and describe the natural $\U(N)$ action carried by the space of $N$ spinors, which will be essential later on. First, we identify the following $\SU(2)$-invariant observables given by the scalar products on the spinors with each other and their dual \cite{un0,un1,un2,un3,un4}:
\be
E_{ij}=\la z_i |z_j\ra,\quad
F_{ij}=[z_i|z_j\ra,\quad
\bF_{ij}=\la z_j |z_i]\,.
\ee
The $E_{ij}$ satisfy $\bE_{ij}=E_{ji}$, while the $F$'s are holomorphic, anti-symmetric and satisfy the Pl\"ucker relations
\be \label{plucker}
F_{ij}F_{kl}=F_{il}F_{kj}+F_{ik}F_{jl}\,.
\ee
The standard scalar products between 3-vectors are easily expressed in terms of these observables:
\be
|\la z_i |z_j\ra|^2
\,=\,
\f12\left(|\vV_i||\vV_j|+\vV_i\cdot\vV_j\right),
\qquad
|[ z_i |z_j\ra|^2
\,=\,
\f12\left(|\vV_i||\vV_j|-\vV_i\cdot\vV_j\right)\,.
\ee
The Poisson brackets of the $E$'s and $F's$ form a closed Lie algebra (for more details see \cite{un2,un3,un4}. In particular, the $E$'s form a $\u(N)$-algebra:
\be
\{E_{ij},E_{kl}\}
\,=\,
-i\left(
\delta_{jk}E_{il}-\delta_{il}E_{kj}
\right)\,.
\ee
Actually, as shown in \cite{un2,un3}, the $E$'s generate the natural $\U(N)$-action on the $N$ spinors: $\{z_i\}\mapsto \{(Uz)_i=\sum_jU_{ij}z_j\}$, for $U\in\U(N)$. The key point is that this action commutes with the closure constraints and is cyclic for fixed total area $A(z_i)\equiv\f12\sum_i \la z_i|z_i\ra$.
Defining the completely squeezed configuration,
\be
\Omega_1=\mat{c}{1 \\ 0},\quad
\Omega_2=\mat{c}{0 \\ 1},\quad
\Omega_{i\ge 3}=0,\quad
A(\Omega_i)=1,
\ee
we can indeed get arbitrary spinors satisfying the closure constraints by acting with unitary matrices on this set of spinors (and appropriately rescaling by the total area):
\be
z_i = A(z_i)\,(U\Omega)_i=A(z_i)\,\mat{c}{U_{i1}\\U_{i2}}\,.
\ee
The closure constraints come from the unitarity of the matrix $U$.

\subsection{Quantization and Coherent Intertwiners}

Following the previous work \cite{un2,un3,un4,spinor,spinor_johannes}, we quantize this classical phase space as a set of $2N$ harmonic oscillators:
\be
z_i^A \arr a_i^A,\quad
\bz_i^A \arr a_i^{A\dagger},\qquad
[a_i^A,a_j^{B\dagger}]=\delta_{ij}\delta^{AB},\quad
[a^A_i,a^B_j]=0\,.
\ee
So now we have a couple of harmonic oscillators $a^{A=0,1}_i$ attached to each leg $i$ of the intertwiner. We then quantize the vectors $\vV_i$ and observables $E_{ij}$, $F_{ij}$ and $\bF_{ij}$, using normal ordering when necessary,
\be
\begin{array}{lcl}
\f12\vV_i= \vsigma^{AB}\bz_i^A z_i^B
&\,\arr\,&
\f12\widehat{\vV}_i=\vsigma^{AB}a_i^{A\dagger} a_i^B
\equiv \vJ_i \\
\f12|V_i|=\f12\la z_i|z_i\ra=\f12\,\bz_i^A z_i^A
&\,\arr\,&
\f12\widehat{|V_i|}= \f12 a_i^{A\dagger} a_i^A
\equiv \cJ_i
\end{array}
\ee
\be
\begin{array}{lcl}
E_{ij}=\la z_i|z_j\ra
&\,\arr\,&
\hE_{ij}=a^{A\dagger}_i a^A_j\\
F_{ij}=[ z_i|z_j\ra
&\,\arr\,&
\hF_{ij}=\epsilon^{AB}a^A_i a^B_j \\
\bar{F}_{ij}=\la z_j|z_i]=-\la z_i|z_j]
&\,\arr\,&
\hFd_{ij}= \epsilon^{AB}a^{A\dag}_i a^{B\dag}_j
\end{array}
\ee
The commutators between these operators reproduce exactly the algebra of the Poisson brackets.
The operators $\vJ_i$ are the generators of the $\su(2)$ algebra attached to the $i$-th leg. Then the total energy on the $i$-th leg, $\cJ_i$, commutes with these generators and give the spin $j_i$ of the $\su(2)$-representation. More precisely, we have Schwinger's representation for the $\su(2)$ algebra and we can easily go between the standard oscillator basis $|n^A_i\ra$ labeled by the number of quanta and the usual magnetic momentum basis $|j_i,m_i\ra$ for spin systems by diagonalizing the operators $J_i^z$ and $\cJ_i$:
\be
|n^0_i,n^1_i\ra_{HO}=|j_i,m_i\ra,
\qquad
\textrm{with}\quad
j_i=\f{n^0_i+n^1_i}2,\quad
m_i=\f{n^0_i-n^1_i}2\,.
\ee
So fixing the total energy of the two harmonic oscillators, we fix the spin $j_i$ of the $\su(2)$-representation attached to the leg $i$.
Calling $\cH^{HO}=\oplus_{n}\C\,|n\ra$ the Hilbert space of a single harmonic oscillator, this allows us to decompose the tensor product $\cH^{HO}\otimes\cH^{HO}$ in $\SU(2)$-representations:
\be
\cH^{HO}\otimes\cH^{HO}=\bigoplus_{j\in\N/2}\cV^j,
\ee
where we write $\cV^j$ for the $\SU(2)$-representation of spin $j$.

We now consider $N$ copies of this representation of $\SU(2)$, and impose the closure constraints $\sum_i \vJ_i$, which amount to require the invariance under the global $\SU(2)$-action. This means that we are looking at $\SU(2)$-invariant states in the tensor product of the $\SU(2)$-representations living on the legs $i$ around the vertex, i.e. \emph{intertwiners} between the spins $j_i$. This defines the Hilbert space of $N$-valent intertwiners from our collection of harmonic oscillators,
\be
\label{int_space}
\cH_N
\,=\,
\textrm{Inv}_{\SU(2)}\bigotimes_i^N (\cH^{HO}_i\otimes\cH^{HO}_i)
\,=\,
\textrm{Inv}_{\SU(2)}\bigotimes_i \bigoplus_{j_i\in\N/2}\cV^{j_i}
\,=\,
\bigoplus_{\{j_i\}} \textrm{Inv}_{\SU(2)} \bigotimes_i \cV^{j_i}.
\ee
The operators $\hE_{ij}$, $\hF_{ij}$ and $\hFd_{ij}$ commutes with the generators of the global $\SU(2)$-transformations, $\sum_i \vJ_i=0$, and thus act on the Hilbert space of intertwiners $\cH_N$. As shown in \cite{un0,un1}, the operators $\hE_{ij}$ form a $\u(N)$-algebra at the quantum level and generate a $\U(N)$ action on intertwiner states, similarly to the $\U(N)$-action on the sets of classical spinors. These $\U(N)$-transformations leave invariant the total area $\cJ\equiv\sum_i \cJ_i=\sum_i \hE_{ii}$. This leads to the following decomposition of the space of $N$-valent intertwiners:
\be
\cH_N
\,=\,
\bigoplus_{\{j_i\in\N/2\}} \textrm{Inv}_{\SU(2)} \bigotimes_i \cV^{j_i}
\,=\,
\bigoplus_{J\in\N} \cR^J\,,\qquad
\cR^J
\,=\,
\bigoplus_{J=\sum_ij_i} \textrm{Inv}_{\SU(2)} \bigotimes_i \cV^{j_i}\,.
\ee
Each subspace $\cR^J$ carries an irreducible representations of $\U(N)$ generated by the operators $\hE_{ij}$ \cite{un1}. Moreover this endows the Hilbert space $\cH_N$ with a Fock space structure, with the operators $\hF_{ij}$ acting as annihilation operators going from $\cR^{J}$ to $\cR^{J-1}$ while the operators $\hFd_{ij}$ act as annihilation operators going from $\cR^{J}$ to $\cR^{J+1}$ \cite{un2}.

\medskip

We can then build coherent states for each of those Hilbert spaces, from the $\SU(2)$ irreducible representations $\cV^j$ to the whole Hilbert space of $N$-valent intertwiners $\cH_{N}$. The coherent intertwiners on $\cH_{N}$ are obtained by group averaging over $\SU(2)$ the harmonic oscillator coherent states. Coherent states on $\cR^J$ and on $\textrm{Inv}_{\SU(2)} \bigotimes_i \cV^{j_i}$ are obtained by projecting them at fixed total area $J$ or at fixed spins $j_i$. Nevertheless, coherent intertwiners were slowly constructed in the reverse order, with a first definition of  the Livine-Speziale coherent intertwiners \cite{ls}, then the definition of the $\U(N)$ coherent states \cite{un2} and finally the introduction of the final coherent intertwiners \cite{un3,un4}. We summarize their definitions and properties below.

\begin{itemize}
\item {\bf $\SU(2)$ Coherent States}:

They are defined by acting with the creation operators $a^{A\dag}$ on the vacuum of the harmonic oscillators, to build the standard coherent states for the harmonic oscillators, and then by projecting to a fixed total energy in order to fix the spin $j$. We denote them $|j,z\ra\in\cV^j$, with a spin label $j$ and a spinor $z\in\C^2$,
\be
|j,z\ra
\,=\,
\f{(z^A a^{A\dag})^{2j}}{\sqrt{(2j)!}}\,|0\ra
\,=\,
\sum_{m=-j}^{+j}\f{\sqrt{(2j)!}}{\sqrt{(j+m)!(j-m)!}}\,(z^0)^{j+m}(z^1)^{j-m}\,|j,m\ra\,.
\ee
Their norm is easy to compute: $\la j,z|j,z\ra=\la z|z\ra^{2j}$. They are coherent states \`a la Perelomov, i.e. they transform covariantly under $\SU(2)$-transformations (e.g. \cite{un2,un3}),
\be
\forall g\in\SU(2),\quad
g\,|j,z\ra
=|j,g\,z\ra\,,
\ee
where $g\in\SU(2)$ acts on the spinor $z$ as a $2\times 2$ matrix in the fundamental $\SU(2)$-representation\footnote{This means that we can generate the $\SU(2)$ coherent states by acting with $\SU(2)$ group elements on the highest weight vector $|j,j\ra$.
Indeed, an arbitrary spinor $z$ can always be obtained from the ``origin spinor" $\Omega=(1,0)$ by a unique $\SU(2)$ transformation:
\be\nn
g(z)\,|\Omega\ra=\f{|z\ra}{\sqrt{\la z|z\ra}},\qquad
g(z)=\f1{\sqrt{\la z|z\ra}}\,\mat{cc}{z^0 & -\bz^1\\ z^1 &\bz^0}
=\f1{\sqrt{\la z|z\ra}}\,(|z\ra,|z])
\ee
This translates into a similar relation on the coherent states,
\be\nn
\f1{\sqrt{\la z|z\ra^{2j}}}|j,z\ra=g(z)\,|j,\Omega\ra=g(z)\,|j,j\ra.
\ee}. Furthermore, these states are the tensor power of the states in the spin-$\f12$ representation, $|j,z\ra=|\f12,z\ra^{\otimes 2j}$.
Finally, these states are semi-classical. They are peaked with minimal uncertainty around the  expectation values of the $\su(2)$-generators $\vJ$:
\be
\f{\la j,z|\vJ|j,z\ra}{\la j,z|j,z\ra}
\,=\,
2j\,\f{\la z|\f{\vsigma}{2}|z\ra}{\la z|z\ra}
\,=\,
j\,\f{\vV(z)}{|\vV(z)|}\,.
\ee

\item {\bf LS Coherent Intertwiners}:

Coherent intertwiners were first introduced in \cite{ls} from tensoring together $N$ $\SU(2)$ coherent states and group-averaging in order to get $\SU(2)$-invariant states. This was re-cast in terms of spinors in \cite{un2,un3}. Such a $N$-valent coherent intertwiner is labeled by a list of $N$ spins $j_i$ and $N$ spinors $z_i$ attached to each leg $i$ and defined by
\be
|\{j_i,z_i\}\ra
\,=\,
\int_{\SU(2)}dg\, g\vartriangleright\bigotimes_i |j_i,z_i\ra
\,=\,
\int_{\SU(2)}dg\, \bigotimes_i g\,|j_i,z_i\ra\,.
\ee
The norm and scalar product of these LS coherent intertwiners can be expressed as a finite sum of ratios of factorials \cite{un2}. Such formulas are also directly deduced from the scalar product of the $\U(N)$ coherent states described below.

An important point is that it is not required that the classical spinors $z_i$ labeling the states satisfy the closure constraint. One can show that the LS coherent intertwiners defined by closed sets of spinors are nevertheless dominant and that those labeled by spinors which do not satisfy the closure are exponentially suppressed \cite{ls}. This is done by computing asymptotically their norm in the large spin regime and showing that closed sets of spinors dominate the integral over coherent states in the decomposition of the identity on the Hilbert space $\textrm{Inv}_{\SU(2)} \bigotimes_i \cV^{j_i}$. Such peakedness properties have been useful to define the EPRL-FK spinfoam models \cite{eprl,fk,ls2}.

As done in \cite{spinor}, it is possible to compute the action of the $\hE$ and $\hF$ operators on these states by commuting their action with the operators $(z^A a^{A\dag})^{2j}$ defining the $\SU(2)$ coherent states\footnote{Notice that the $\hE$ and $\hF$ operators are $\SU(2)$-invariant and thus commute with the group averaging}. This gives
\be
\begin{aligned}
\hE_{ij}\,|\{j_k,z_k\}\ra &=
\f{\sqrt{2j_j}}{\sqrt{2j_i+1}}\,
\left(z_j^A\f{\pp}{\pp z_i^A}\right)\,|\{j_i+\f12,j_j-\f12,j_k,z_k\}\ra,\\
\hF_{ij}\,|\{j_k,z_k\}\ra &=
\sqrt{(2j_i)(2j_j)}\,
F_{ij}\,|\{j_i-\f12,j_j-\f12,j_k,z_k\}\ra,\\
\hFd_{ij}\,|\{j_k,z_k\}\ra &=
\f{1}{\sqrt{(2j_i+1)(2j_j+1)}}\,
\left(\eps^{AB}\f{\pp^2}{\pp z_i^A\pp z_j^B}\right)
\,|\{j_i+\f12,j_j+\f12,j_k,z_k\}\ra,
\end{aligned}
\ee

\item {\bf $\U(N)$ Coherent States}:

They are defined on $\cR^J$ for fixed total area $J=\sum_i j_i$, \cite{un2},
\be
|J,\{z_i\}\ra
\,=\,
\f{1}{\sqrt{J!(J+1)!}}\,
\biggl(\f12\sum_{i,j}[z_i|z_j\ra\,\hFd_{ij}\biggr)^J\,|0\ra\,.
\ee
They are superpositions of LS coherent intertwiners \cite{un2} as follows
\be
\f{1}{\sqrt{J!(J+1)!}}\,|J,\{z_i\}\ra
\,=\,
\sum_{J=\sum_i j_i}\f{1}{\sqrt{\prod_i (2j_i)!}}\,|\{j_i,z_i\}\ra
\,=\,
\f{1}{(2J)!}\int dg\,g\rhd(\sum_i z_i^Aa^{A\dag}_i)^{2J}\,|0\ra\,.
\ee
From their definition above, one can prove that they are covariant under the $\U(N)$-action \cite{un2}, hence the name of $\U(N)$ coherent states,
\be
\hat{U}\,|J,\{z_i\}\ra
\,=\,
|J,\{(Uz)_i\}\ra,\qquad
U=e^{i\alpha},
\quad
\hat{U}=e^{i\sum_{i,j}\alpha_{ij}\hE_{ij}}\,,
\ee
where the arbitrary $N\times N$ Hermitian matrix $\alpha$ generates the unitary $\U(N)$ transformation.
Their scalar products and norms are explicitly known \cite{un2},
\beq
\label{un_scalar}
\la J,\{w_i\}|J,\{z_j\}\ra
&=&
\det\,\biggl(\sum_i |z_i\ra\la w_i|\biggr)^J
=
\biggl(\f12\sum_{i,j} \la w_j|w_i][ z_i|z_j\ra\biggr)^J
=
\biggl(\f12\sum_{i,j} \bF_{ij}(w)F_{ij}(z)\biggr)^J
,
\\
\la J,\{z_i\}|J,\{z_i\}\ra
&=&
\f1{2^{2J}}\biggl[
\Bigl(\sum_i \la z_i|z_i\ra\Bigr)^2
-\Bigl(\sum_i \la z_i|\vsigma|z_i\ra\Bigr)^2
\biggr]^{J}
\,=\,
\f1{2^{2J}}\biggl[
\Bigl(\sum_i |\vV_i|\Bigr)^2
-\Bigl|\sum_i \vV_i\Bigr|^2
\biggr]^{J}\,.
\eeq
When the closure constraints are satisfied, i.e. when $\sum_i |z_i\ra\la z_i|\propto \id$, the norm simplifies to $\la J,\{z_i\}|J,\{z_i\}\ra=A(z_i)^{2J}$ where $A$ is the total area $A(z)=\f12\sum_i\la z_i|z_i\ra$.
The action of the operators $\hE$ and $\hF$ reads \cite{un3,spinor}:
\be
\begin{aligned}
\hE_{ij}\,|J,\{z_k\}\ra &=
\left(z_j^A\f{\pp}{\pp z_i^A}\right)\,|J,\{z_k\}\ra,\\
\hF_{ij}\,|J,\{z_k\}\ra &=
\sqrt{J(J+1)}\,F_{ij}\,|J-1,\{z_k\}\ra,\\
\hFd_{ij}\,|J,\{z_k\}\ra &=
\f1{\sqrt{(J+1)(J+2)}}\,\left(\eps^{AB}\f{\pp^2}{\pp z_i^A\pp z_j^B}\right)
\,|J+1,\{z_k\}\ra,
\end{aligned}
\ee
Finally, all these properties allow to compute exactly the expectation values of the $\hE$ operators \cite{un2}:
\be
\f{\la J,\{z_i\}|\hE_{ij}\,|J,\{z_i\}\ra}{\la J,\{z_i\}|J,\{z_i\}\ra}
=
J\,\f{\la z_i|z_j\ra}{\f12\sum_k\la z_k|z_k\ra}
=
J\,\f{E_{ij}}{A},
\ee
where we assumed that the spinors $z_i$ satisfy the closure condition. Let us emphasize that this expectation value is \emph{exact} while the expectation values of the $\SU(2)$-observables on the LS coherent intertwiners are only known asymptotically in the large spin limit.

\item {\bf Coherent Intertwiners}:

The last notion of coherent intertwiners was introduced in \cite{un3}. They truly represent coherent states on the spinorial phase space: they are simply labeled by a phase space point, i.e. $N$ spinors (up to global $\SU(2)$ rotations). More explicitly, they are defined as the eigenstates of the annihilation operators $\hF_{ij}$ (which is possible since the operators $\hF_{ij}$ all commute with each other). Their expansions in the previous bases are
\be
|\{z_i\}\ra
\,=\,
\sum_J\f{1}{\sqrt{J!(J+1)!}}\,|J,\{z_i\}\ra
\,=\,
\sum_{\{j_i\}}\f{1}{\prod_i\sqrt{(2j_i)!}}\,|\{j_i,z_i\}\ra
\,=\,
\int dg\,g\vartriangleright
e^{\sum_iz_i^A a_i^{A\dag}}\,|0\ra,
\ee
The last equality shows that these coherent intertwiners $|\{z_i\}\ra$ are the group averaging of the standard (unnormalized) coherent states for the harmonic oscillators. Using the above expansion onto the states $|J,\{z_i\}\ra$ and the action of the annihilation operators on them\footnote{
Since the operator $\hF_{ij}$ is $\SU(2)$-invariant and thus commute with the $\SU(2)$-action, we could more simply compute its commutator with the usual operator $e^{\sum_k z_k^A a_k^{A\dag}}$. Indeed, we get the same results by computing
\be
\Big{[}\hF_{ij},\sum_kz_k^A a_k^{A\dag}\Big{]}
\,=\,
\epsilon^{AB}(z_i^Aa^B_j-z^A_ja^B_i),
\qquad
\Big{[}\epsilon^{AB}(z_i^Aa^B_j-z^A_ja^B_i),\sum_kz_k^C a_k^{C\dag}\Big{]}
\,=\,
2F_{ij}\,\id\,.
\ee
}, we easily show that
\be \label{hF_CI}
\hF_{ij}\,|\{z_k\}\ra
\,=\,
[z_i|z_j\ra\,|\{z_k\}\ra
\,=\,
F_{ij}\,|\{z_k\}\ra\,.
\ee
We can similarly compute the action of the other $\SU(2)$-invariant operators $\hE_{ij}$ and $\hFd_{ij}$,
\be
\hE_{ij}\,|\{z_k\}\ra =
\left(z_j^A\f{\pp}{\pp z_i^A}\right)\,|\{z_k\}\ra,\qquad
\hFd_{ij}\,|\{z_k\}\ra =
\left(\eps^{AB}\f{\pp^2}{\pp z_i^A\pp z_j^B}\right)
\,|\{z_k\}\ra.
\ee
We further compute the norm and scalar product of these states \cite{un3}:
\beq
\la\{w_i\}|\{z_i\}\ra
&=&
\sum_J\f{1}{J!(J+1)!}\,
\la J,\{w_i\}|J,\{z_i\}\ra
=\sum_J\f{1}{J!(J+1)!}\,
\left(\det \sum_i |z_i\ra\la w_i|\right)^J\\
\la\{z_i\}|\{z_i\}\ra&=&\sum_J\f{A(z)^{2J}}{J!(J+1)!}
=\f{I_1(2A(z))}{A(z)}
\quad\textrm{assuming the closure constraint on the } z_i,
\eeq
where the $I_n$ are the modified Bessel functions of the first kind.
Finally, we also give the expectation values of the $\hE$-operators:
\be
\label{Evalue}
\la\{z_k\}|\hE_{ij}\,|\{z_k\}\ra
\,=\,
\f{E_{ij}}{A(z)}\sum_{J\ge 1} \f{(A(z))^{2J}}{(J-1)!(J+1)!}
\,=\,
\f{E_{ij}}{A(z)}\,I_2(2A(z)).
\ee
The asymptotic behavior of the $| \{z_e\}\ra$ coherent states  for large area $A(z)\gg 1$ is given by\footnote{
The asymptotic for the modified Bessel functions $I_n(x)$ for $x\in\R$ do not depend on the label $n$ at leading order:
\be\nn
I_n(x)\,\underset{x\arr+\infty}{\sim}\,\f{e^x}{\sqrt{2\pi x}}\,\left(1+\cO\left(\f1x\right)\right).
\ee
}
\be
\la \{z_k\}|\{z_k\}\ra
\,\sim\,
\f{e^{2A(z)}}{\sqrt{4\pi}\,A(z)^{3/2}} \,,\qquad
\f{\la\{z_k\}|\hE_{ij}\,|\{z_k\}\ra}{\la \{z_k\}|\{z_k\}\ra}
\,\sim\,
E_{ij}
\,,
\ee
showing that these coherent intertwiners have the right semi-classical behavior.
\end{itemize}

\subsection{From spinors to SU(2) invariant variables} \label{sec:SU2 invariant}

In this article it will be convenient to work sometimes with $\SU(2)$ invariant variables instead of spinors. One way to get them is is as follows. First we consider the variables $F_{ij}=[z_i|z_i\rangle$ formed from $z_i$. But they are not independent variables, due to the Pl\"u{}cker relations \eqref{plucker}. The latter exhaust the dependence relations between the $F_{ij}$, and can be solved to extract the $\SU(2)$ invariant content. One uses the Pl\"u{}cker relations to express some of the $F_{ij}$ in terms of others. Depending on which of them we eliminate, one gets different sets of variables. For instance, one can choose
\be \label{su2 invariant var}
\{ z_i\}\ \longrightarrow \ \{ F_{12}, F_{13}, F_{23},\dotsc,F_{N3}, Z_4,\dotsc,Z_N\}\,,
\ee
where the variables $Z_k$, for $k=4,\dotsc,N$ are the cross-ratios,
\be \label{def cross-ratio}
Z_k = \frac{F_{k1}\,F_{23}}{F_{k3}\,F_{12}}\,.
\ee
We will show how that is done in practice in the section \ref{sec:direct expansion}.

That gives $2(N-3)+3$ (complex) variables per intertwiner, which is the expected counting in agreement with the standard $3$-valent tree unfolding ($N$ spins plus $N-3$ internal spins) of $\SU(2)$ intertwiners. The choice of which $F_{ij}$ are eliminated corresponds to a choice of cross-ratios. We expect that choice to be equivalent to the choice of a tree $\cT^\beta$ to unfold the intertwiner, as already shown in \cite{holquantumtet} for $N=4$.

Then, one can try to use those variables to build coherent intertwiners. The simplest way is to re-express the coherent intertwiners described above. It can be done as in \cite{holquantumtet} by studying the action of $\SL(2,\C)$ on spinors to extract the dependence on the $F_{ij}$. One gets for LS intertwiners
\be \label{factor coherent state}
|\{j_i,z_i\}\ra = F_{12}^{J-2j_3} F_{13}^{2j_1+2j_3-J} F_{23}^{2j_2+2j_3-J} \prod_{k=4}^N F_{k3}^{2j_k}\ |\{j_i, Z_k\}\ra\,,
\ee
where $|\{j_i,Z_k\}\ra$ is a state which only depends on the cross-ratios, and $J = \sum_{i=1}^N j_i$. Other equivalent choices of factorization can be obtained by acting with elements of the permutation group to exchange some links \cite{holquantumtet}.

A direct derivation of \eqref{factor coherent state} via the $\U(N)$ scalar product and the Pl\"{u}cker identities will be given in the section \ref{sec:direct expansion}.

The scalar product between the states $|\{j_i,Z_k\}\ra$ is known in the case $N=4$, \cite{holquantumtet} but not in general\footnote{Note that the case $N=3$ is somewhat trivial as there is no cross-ratio.}. In this paper we will show a generic formula for this scalar product. It has the following polynomial form which is different of that of \cite{holquantumtet} for $N=4$.
\begin{res} \label{res:scalarprod}
The scalar product $\la \{j_i,Z_k\}|\{j_i,W_k\}\ra$ admits the form
\begin{multline}
\f{(1+\sum_i j_i)!}{\prod_i (2j_i)!}\ \la \{j_i,Z_k\}|\{j_i,W_k\}\ra
\\=
\sum_{\{p_{1k},p_{2k},p_{kl}\}_{k\geq4, l>k}} \frac{1}{\displaystyle{\bigl(J-2j_3-\sum_{k\geq 4}(p_{1k}+p_{2k}) - \sum_{4\leq k<l}p_{kl}\bigr)! \bigl(2j_1+2j_3-J+\sum_{k\geq4}p_{2k}+\sum_{4\leq k<l}p_{kl}\bigr)!}}\\
\frac{\prod_{k=4}^N [\bar{Z}_k W_k]^{p_{1k}}\,[(1+\bar{Z}_k)(1+W_k)]^{p_{2k}}\,\prod_{l>k\geq 4}\Bigl[\bigl(\bar{Z}_k-\bar{Z}_l\bigr)\bigl(W_k-W_l\bigr)\Bigr]^{p_{kl}}}{\displaystyle{\bigl(2j_2+2j_3-J+\sum_{k\geq4}p_{1k}+\sum_{4\leq k<l}p_{kl}\bigr)!\prod_{k\geq4}\bigl(2j_k-p_{1k}-p_{2k}-\sum_{l\neq k\geq4}p_{kl}\bigr)! p_{1k}! p_{2k}! \prod_{4\leq k<l}p_{kl}!}}\,.
\end{multline}
\end{res}
We will show this result in the section \ref{sec:direct expansion} and also build generating functions for these scalar products.

\section{Evaluations of Coherent Spin Networks as Generating Functions} \label{sec:GF-2vertex}

\subsection{Coherent Spin Networks on the 2-Vertex Graph}

\subsubsection{Classical Phase Space on the 2-Vertex Graph}

\begin{figure}[h]
\begin{center}
\includegraphics[height=40mm]{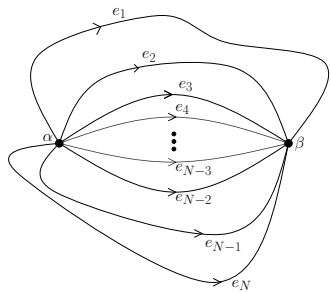}
\caption{The 2-vertex graph: the two vertices $\alpha$ and $\beta$
are linked by $N$ edges. We choose all the edges oriented in the same way with $\alpha$ as their source vertex while $\beta$ is the target vertex. The spinors $z_i$ live around the vertex $\alpha$ while the spinors $w_i$ live on the legs of the vertex $\beta$. {\it By courtesy of I\~naki Garay from \cite{spinor}.} \label{2vertex}}
\end{center}
\end{figure}

Let us consider the 2-vertex graph, made of two vertices $\alpha$ and $\beta$ connected by $N$ links, as pictured on the figure \ref{2vertex}, and start by describing the classical phase space of spinors on that graph, as defined in \cite{un4,spinor,spinor_johannes,sfcosmo}.
We have two sets of spinors, $z_i$ at the vertex $\alpha$ and $w_i$ at the vertex $\beta$, both satisfying the closure constraints, $\sum_i |z_i\ra\la z_i|\propto\id$ and $\sum_i |w_i\ra\la w_i|\propto\id$, which translates into $\sum_i \vV(z_i)=\sum_i \vV(w_i)=0$ in terms of 3-vectors.
Moreover, we impose matching conditions, $\la z_i|z_i\ra=\la w_i|w_i\ra$ or equivalently $|\vV(z_i)|=|\vV(w_i)|$ for all edges $i$. All these constraints form a first class system. While the closure constraints generate $\SU(2)$ transformations at each vertex, the matching conditions generate $\U(1)$-phase multiplications on the spinors on each edge $i$. The resulting phase space on the 2-vertex graph is then defined as the symplectic reduction $\C^{4N}//(\SU(2)^2\times\U(1)^N)$. This constrained phase space has dimension $2\times(3N-6)$ and can be identified with the gauge invariant holonomy-flux phase space of loop quantum gravity on the 2-vertex graph, for which the configuration space is defined as the set of group elements $g_1,..,g_N\in\SU(2)$ up to global left and right $\SU(2)$ translations, $g_i\arr g^{-1}g_ih$. Furthermore, the space of holomorphic $L^2$ functions on $\C^{4N}//(\SU(2)^2\times\U(1)^N)$ is shown to be isomorphic to the space of $L^2$-functions on $\SU(2)^N/\SU(2)^2$ i.e. to the Hilbert space  of spin network functions on the 2-vertex graph \cite{spinor,spinor_johannes}.

This isomorphism is realized through the reconstruction of the holonomies $g_i\in\SU(2)$ from the spinor variables as first shown in \cite{twisted2} and further investigated in \cite{spinor,spinor_johannes}:
\be
\label{su2_spinor}
g_i=\f{|z_i]\la w_i|-|z_i\ra[ w_i|}{\sqrt{\la w_i|w_i\ra\la z_i|z_i\ra}}
\,\in\SU(2),
\qquad
g_i|w_i\ra\,=\, |z_i],
\quad
g_i|w_i]\,=\, -|z_i\ra\,.
\ee
Indeed, when we assume the matching condition, i.e. that the spinors have equal norm, $\la z_i|z_i\ra=\la w_i|w_i\ra$, this is the unique $\SU(2)$ group element $g_i$ mapping the spinor $w_i$ on $z_i$.

A more detailed analysis of the classical phase space associated to the 2-vertex graph, its various parameterizations, its geometrical interpretation and its relevance for defining cosmological settings in loop (quantum) gravity can be found in \cite{sfcosmo}.

\medskip

Flat configurations are defined by $g_i=\id,\,\forall i$ up to $\SU(2)$ gauge transformations at the two vertices, i.e. $g_i$ all equal to one fixed group element $g\in\SU(2)$ for all edges $i$. This group element $g$ maps all the spinors $|w_i\ra$ on their counterpart $|z_i]$. In particular, it implies that the $\SU(2)$-invariant observables $E$ and $F$ are equal for both sets of spinors $w_i$ and $\varsigma z_i$, i.e. $E_{ij}(z)=\bE_{ij}(w)$ and $F_{ij}(z)=\bF_{ij}(w)$.

We can actually go further. Indeed, as shown by proposition 1.2 in \cite{un4}, the $F_{ij}$ are a complete set of $\SU(2)$-invariant observables. Assuming the closure constraints on both sets of spinors $w_i$ and $z_i$, then assuming $F_{ij}(z)=\bF_{ij}(w)$, i.e. $[z_i|z_j\ra=-\la w_i|w_j]$, for all pairs of (different) edges is equivalent to the existence of a group element $g\in\SU(2)$ such that $|z_i]=g|w_i\ra$ for all edges $i$:
\be \label{flat-constraint}
F_{ij}(z)=\bF_{ij}(w),\,\,\forall i,j
\quad
\Longleftrightarrow
\quad
\exists g\in\SU(2), \,|z_i]=g|w_i\ra\,\,\forall i.
\ee
This fully characterizes the flat configurations on the 2-vertex graph.

\subsubsection{Quantum States}

Quantum states of geometry on this graph, as defined by loop quantum gravity, are gauge invariant  functions of the holonomies along its edges, i.e. functions $\vphi(g_1,..,g_N)$ of $N$ group elements with a $\SU(2)$ invariance at both vertices:
\be
\vphi(g_1,..,g_N)=\vphi(g^{-1}g_1h,..,g^{-1}g_Nh),
\qquad\forall g_i,g,h\in\SU(2)\,.
\ee
We endow the space of such functions with the Haar measure on $\SU(2)^N$, which allows to define
the Hilbert space of quantum states $L^2(\SU(2)^N/\SU(2)^2)$ with the natural scalar product:
\be
\la \vphi|\tilde{\vphi}\ra
\,=\,
\int_{\SU(2)^N} [dg_i]^N\,
\overline{\vphi(g_1,..,g_N)}\,\tilde{\vphi}(g_1,..,g_N)\,.
\ee
Using the Peter-Weyl theorem to decompose $L^2$ functions on $\SU(2)$ in Wigner matrices, one can show that a basis of this Hilbert space is provided by the spin network states. A spin network is defined by assigning a $\SU(2)$-representation $j_i\in\N/2$ to each edge of the graph and choosing two intertwiners $i_\alpha$ and $i_\beta$ in $\textrm{Inv}_{\SU(2)} \bigotimes_i \cV^{j_i}$ for the two vertices. Then the spin network function reads
\be
\vphi_{j_i,i_\alpha,i_\beta}(g_1,..,g_N)
\,=\,
\sum_{a_i,b_i}
\prod_i \overline{i_\alpha^{a_1..a_N}}\,D^{j_i}_{a_ib_i}(g_i)\,i_\beta^{b_1..b_N}
\,\equiv\,
\la j_1,..,j_N,i_\alpha,i_\beta| g_1,..,g_N\ra\,.
\ee
The scalar  product between these functions is easily computed and is actually equal to the scalar product on the intertwiner space,
\be
\la \vphi_{j_i,i_\alpha,i_\beta}|\vphi_{\tilde{j}_i,\tilde{i}_\alpha,\tilde{i}_\beta}\ra
\,=\,
\prod_i \f{1}{2j_i+1}\delta_{j_i\tilde{j}_i}\,
\la \tilde{i}_\alpha|i_\alpha\ra\,\la i_\beta| \tilde{i}_\beta\ra\,.
\ee

We can now define coherent spin network states on the 2-vertex graph by assigning coherent intertwiners to both vertices $\alpha$ and $\beta$. Choosing spinors $w_i$ and $z_i$, we consider the corresponding coherent intertwiners $|\{w_i\}\ra$ and $|\{z_i\}\ra$ and define the following coherent spin network functions \cite{un4,un4_conf}:
\be
\vphi_{w_i,z_i}(g_i)
\,\equiv\,
\la \{\varsigma z_i\}|\otimes_i g_i|\{w_i\}\ra
\,=\,
\int_{\SU(2)^2} dg\,dh\
e^{\sum_i [z_i|gg_ih|w_i\ra}\,.
\ee
Expanding coherent intertwiners into LS intertwiners, we can decompose this coherent spin network function in standard spin network states with fixed spins $j_i$ on the edges,
\be
\vphi_{w_i,z_i}(g_i)
\,=\,
\sum_J
\f1{J!(J+1)!}\la J,\{\varsigma z_i\}|\otimes g_i|J,\{w_i\}\ra
\,=\,
\sum_{\{j_i\}}\f1{\prod_i (2j_i)!}\,\la \{j_i,\varsigma z_i\}|\otimes g_i|\{j_i,w_i\}\ra\,.
\ee

\subsection{Generating function for scalar products of coherent intertwiners} \label{sec:direct expansion}

We now evaluate these functions at the identity,
\be \label{eval coherent spinnet}
\cW(w_i,z_i)\equiv\vphi_{w_i,z_i}(\id)
=
\sum_{\{j_i\}}\f1{\prod_i (2j_i)!}\,\la \{j_i,\varsigma z_i\}|\{j_i,w_i\}\ra\,.
\ee
These objects can be interpreted as generating functions. To see that we recourse to the $\SU(2)$ invariant variables proposed in the section \ref{sec:SU2 invariant}. Using the factorization \eqref{factor coherent state} (and because the cross-ratios for the variables $\varsigma z_i$ are the complex conjugate $\bar Z_k$), we get
\begin{multline} \label{GF scalarprod}
\cW = \sum_{\{j_i\}}
[F_{12}(z)F_{12}(w)]^{J-2j_3} [F_{13}(z)F_{13}(w)]^{2j_1+2j_3-J} [F_{23}(z)F_{23}(w)]^{2j_2 +2j_3-J} \prod_{k=4}^N [F_{k3}(z) F_{k3}(w)]^{2j_k}\\
\times \ \f1{\prod_i (2j_i)!}\ \la \{j_i,\bar{Z}_k\}|\{j_i,W_k\}\ra\,.
\end{multline}
This shows that $\cW$ is a generating function for the scalar product $\la \{j_i,\bar{Z}_k\}|\{j_i,W_k\}\ra$.

Remarkably, $\cW$ admits another expansion that leads to the result \ref{res:scalarprod} from the previous section. Using \eqref{un_scalar} for the scalar product of $\U(N)$ coherent states into \eqref{eval coherent spinnet}, one gets
\be
\cW(w_i,z_i)
\,=\,
\sum_J
\f{1}{J!(J+1)!}
\bigl(\det_{2\times 2} X\bigr)^J\,,
\ee
where $X$ is a $2\times 2$ matrix, $X=\sum_i |w_i\ra [z_i|$ whose determinant is simply evaluated
\be
\det X = \sum_{i<j} F_{ij}(z)F_{ij}(w)\,.
\ee
Then, the multinomial expansion yields an expression in which the natural variables are $\{F_{ij}\}$,
\be
\cW(w_i,z_i)
\,=\,
\sum_{\{p_{ij}\}} \frac{1}{(1+\sum_{i<j}p_{ij})!} \prod_{i<j} \frac{1}{p_{ij}!}\Bigl[F_{ij}(z)\,F_{ij}(w)\Bigr]^{p_{ij}}\,.
\ee
However the $F_{ij}$ are not independent. First, we re-arrange their product,
\be
\prod_{i<j} F_{ij}(z)^{p_{ij}} = F_{12}^{p_{12}}\, F_{13}^{p_{13}}\, F_{23}^{p_{23}}\, \prod_{i=4}^N F_{1i}^{p_{1i}}\, F_{2i}^{p_{2i}}\, F_{3i}^{p_{3i}} \prod_{4\leq k<l} F_{kl}^{p_{kl}}\,.
\ee
To see how the cross-ratios \eqref{def cross-ratio} appear, we use the Pl\"{u}cker relations appropriately. The variables $F_{2i}$ are related by $F_{2i} F_{1j} = F_{12} F_{ji} - F_{1i} F_{j2}$. Choosing $j=3$, one gets for all $i\geq4$
\be \label{plucker F2i}
F_{2i} = \frac{F_{12}\, F_{3i}}{F_{13}}\ \bigl(1+Z_i\bigr)\,.
\ee
Then one uses the definition of the cross-ratio $Z_i$, for all $i\geq 4$, to rewrite $F_{1i}$ like $F_{1i} = \frac{F_{12}\,F_{3i}}{F_{23}}\ Z_i$. Finally, Pl\"{u}cker relations are used to eliminate $F_{kl}$, $4\leq k<l$,
\be \label{plucker Fkl}
F_{kl} = \frac{F_{1k}\,F_{3l} + F_{1l}\,F_{k3}}{F_{13}} = \frac{F_{12}}{F_{13}\,F_{23}}\, F_{3k}\,F_{3l}\,\bigl(Z_k-Z_l\bigr)\,.
\ee
Gathering those pieces,
\begin{multline}
\prod_{i<j} F_{ij}(z)^{p_{ij}} = F_{12}^{p_{12}+\sum_{i=4}^N(p_{1i}+p_{2i})+\sum_{l>k\geq4} p_{kl}}\ F_{13}^{p_{13}-\sum_{i=4}^N p_{2i}-\sum_{l>k\geq4} p_{kl}}\ F_{23}^{p_{23}-\sum_{i=4}^N p_{1i}-\sum_{l>k\geq4} p_{kl}}\\
\biggl[\,\prod_{i=4}^N F_{3i}^{p_{1i}+p_{2i}+p_{3i}}\,\prod_{l>k\geq 4} \bigl(F_{3k}\,F_{3l}\bigr)^{p_{kl}}\biggr]\,
\prod_{i=4}^N Z_i^{p_{1i}}\,(1+Z_i)^{p_{2i}}\,\prod_{l>k\geq 4}\bigl(Z_k-Z_l\bigr)^{p_{kl}}\,.
\end{multline}
The $N(N-1)/2$ integers $\{p_{ij}\}$ are summed over. We now make a change of variables on $N$ of them,
\be
\{p_{ij}\}_{1\leq i<j\leq N}\ \longrightarrow\ \Bigl\{j_i = \frac12 \sum_{j\neq i}p_{ij}\Bigr\}_{i=1,\dotsc,N}, \{p_{1k},p_{2k}\}_{k=4,\dotsc,N}, \{p_{kl}\}_{4\leq k<l\leq N}\,.
\ee
The spins $j_i$ satisfy $\sum_{i=1}^N j_i = \sum_{i<j}p_{ij} = J$, which implies that once they are fixed, the $N(N-3)/2$ remaining variables $\{p_{1k},p_{2k},p_{kl}\}$ are bounded. Finally, we obtain
\be
\prod_{i<j} F_{ij}(z)^{p_{ij}} = F_{12}^{J-2j_3}\, F_{13}^{2j_1+2j_3-J}\, F_{23}^{2j_2 +2j_3-J}\, \prod_{i=4}^N F_{3i}^{2j_i}\ \prod_{k=4}^N Z_k^{p_{1k}}\,(1+Z_k)^{p_{2k}}\,\prod_{l>k\geq 4}\bigl(Z_k-Z_l\bigr)^{p_{kl}}\,.
\ee

Applying the same reasoning to the two nodes of the graph, one arrives at
\begin{multline} \label{GF}
\cW = \sum_{\{j_i,p_{1k},p_{2k},p_{kl}\}} \frac1{(1+\sum_i j_i)!} \frac{[F_{12}(z)F_{12}(w)]^{J-2j_3} [F_{13}(z)F_{13}(w)]^{2j_1+2j_3-J} [F_{23}(z)F_{23}(w)]^{2j_2 +2j_3-J}}{\displaystyle{\bigl(J-2j_3-\sum_{k\geq 4}(p_{1k}+p_{2k}) - \sum_{4\leq k<l}p_{kl}\bigr)! \bigl(2j_1+2j_3-J+\sum_{k\geq4}p_{2k}+\sum_{4\leq k<l}p_{kl}\bigr)!}}\\
\frac{\prod_{i=4}^N [F_{3i}(z)F_{3i}(w)]^{2j_i}\ \prod_{k=4}^N [Z_k W_k]^{p_{1k}}\,[(1+Z_k)(1+W_k)]^{p_{2k}}\,\prod_{l>k\geq 4}\Bigl[\bigl(Z_k-Z_l\bigr)\bigl(W_k-W_l\bigr)\Bigr]^{p_{kl}}}{\displaystyle{\bigl(2j_2+2j_3-J+\sum_{k\geq4}p_{1k}+\sum_{4\leq k<l}p_{kl}\bigr)!\prod_{k\geq4}\bigl(2j_k-p_{1k}-p_{2k}-\sum_{l\neq k\geq4}p_{kl}\bigr)! p_{1k}! p_{2k}! \prod_{4\leq k<l}p_{kl}!}}\,.
\end{multline}
The result \ref{res:scalarprod} follows from comparing the above expansion with the expansion \eqref{GF scalarprod}.


\subsection{More Generating Functions for Intertwiner scalar products}

The interpretation of the coherent spin network as a generating function suggests a natural generalization which consists in changing the factorial weights in the sums. We will not consider arbitrary factors depending on the individual spins $j_i$ but will restrict ourselves to global factors depending on the total area $J=\sum_i j_i$. This amounts to changing the normalization of the $\U(N)$ intertwiners in the definition of the spin network function. Considering a sequence $f_J\in\C$, we define the modified spin network function as
\be \label{modified def}
\vphi_{w_i,z_i}^{f_J}(g_i)
\,=\,
\sum_J
\f{f_J}{J!(J+1)!}\la J,\{\varsigma z_i\}|\otimes g_i|J,\{w_i\}\ra
\,=\,
\sum_{\{j_i\}}
\f{f_{\sum_i j_i}}{\prod_i {(2j_i)!}}\,\la \{j_i,\varsigma z_i\}|\otimes g_i|\{j_i,w_i\}\ra\,,
\ee
and the corresponding evaluation
\be
\cW^{f_J}(w_i,z_i)
\,=\,
\sum_J
\f{f_J}{J!(J+1)!}\la J,\{\varsigma z_i\}|J,\{w_i\}\ra
\,=\,
\sum_{\{j_i\}}
\f{f_{\sum_i j_i}}{\prod_i {(2j_i)!}}\,\la \{j_i,\varsigma z_i\}|\{j_i,w_i\}\ra\,.
\ee
Different choices of coefficients $f_J$ can then be interpreted as different choices of generating functions for the scalar product of coherent intertwiners.
The coherent spin network case of the previous section obviously corresponds to $f_J=1$. The choice $f_J=(J+1)!^2$ has been considered on several occasions in the literature, in the context of generating functions for Wigner 3nj-symbols. It started with the seminal work of Schwinger \cite{schwinger:52}, who found the generating functions of 6j-symbols and 9j-symbols, using the bosonic creation and annihilation operators. His work was revisited later by Bargmann \cite{bargmann:62} using Gaussian integrals. Bargmann's approach was then generalized to 12j-symbols and 15j-symbols of the first kind (and potentially to all 3nj-symbols of the first kind). The generic structure of generating functions for any 3nj-symbols has regularly attracted several authors who found remarkable closed formulas \cite{labarthe, schnetz}. More recently, mathematicians have been revisiting 3nj-symbols as trivalent spin network evaluations \cite{garoufalidis, costantino-generating}. From the physics point of view, both choices $f_J=1$ and $f_J=(J+1)!^2$ have been considered to express the dynamics of coherent spin networks in the topological BF model, for trivalent graphs, in \cite{recursion_spinor}.

Remarkably, most of these works use methods and variables similar to those which have been recently introduced in the context of loop quantum gravity, like the spinors and their brackets $F_{ij}$ which are used \cite{costantino-generating}. However, the difference with the present work and loop quantum gravity interests is that we are not only interested in trivalent spin network graphs, but more generally in graphs with nodes of arbitrary degrees. During the completion of this work, the task of finding generating functions in the presence of $4$-valent nodes has been carried out by Freidel and Hnybida in \cite{jeff}.




%
%

As in the case $f^J=1$, the formula \eqref{un_scalar} for the scalar product of $\U(N)$ coherent states can be used to compute the generating functions as
\be
\cW^{f_J}(w_i,z_i)
\,=\,
\sum_J
\f{f_J}{J!(J+1)!}
(\det_{2\times 2} X)^J,
\ee
with $X=\sum_i |w_i\ra [z_i|$. Depending on the coefficients $f_J$, the convergence properties of the series change. Let us look at the main cases\footnote{
Due to the specific changes of the weights in the series' coefficients by $(J+1)!$ factors, these generating functions are actually almost the Borel transforms of one another. For instance, comparing the exponential and algebraic functions, we have:
\be
\sum_J \f{1}{J!}x^J= e^{x}
\,\Rightarrow\,
\sum_J \f{(J+1)}{J!}x^J= \pp_x(xe^{x})=(x+1)e^x
\,\Rightarrow\,
\sum_J (J+1)x^J=\sum_J J!\,\f{(J+1)}{J!}x^J
=\int_0^{+\infty}dt\,e^{-t}(tx+1)e^{tx}
=\f1{(1-x)^2}\,.
\ee
}
\beq
f_J=1 & \quad\longrightarrow\quad &
\cW(w_i,z_i)= \sum_J \f{1}{J!(J+1)!}(\det X)^J
= \f{I_1(2\sqrt{\det X})}{\sqrt{\det X}} \label{coh}\\
f_J=(J+1)! &\quad\longrightarrow\quad  &
\cW^{exp}(w_i,z_i)= \sum_J \f{1}{J!}(\det X)^J
= e^{\det X} \label{exp}\\
f_J=(J+1)!^2 & \quad\longrightarrow\quad  &
\cW^{alg}(w_i,z_i)= \sum_J (J+1)(\det X)^J
= \f1{(1-\det X)^2}\,.\label{alg}
\eeq
The coherent spin network evaluation $\cW(w_i,z_i)$ and the exponential generating function $\cW^{exp}(w_i,z_i)$ always converge. On the other hand, the algebraic generating function $\cW^{alg}(w_i,z_i)$, with $f^J=(J+1)!^2$ as introduced by Schwinger, has a pole at $\det X=1$, but has the advantage of being a simple rational function in the spinor variables $z_i$ and $w_i$.

It is also possible to recast them as functions of the $\SU(2)$ invariant variables \eqref{su2 invariant var}. We only have to re-express $\det X$ appropriately, using the Pl\"ucker relations like in \eqref{plucker F2i}, \eqref{plucker Fkl},
\begin{multline}
\det X = F_{12}(z)F_{12}(w)+ F_{13}(z)F_{13}(w)+F_{23}(z)F_{23}(w)\\
+ \sum_{k\geq4} F_{3k}(z)F_{3k}(w) \biggl[1+ F_{12}(z)F_{12}(w)
\biggl(\frac{Z_k W_k}{F_{23}(z)F_{23}(w)} + \frac{(1+Z_k)(1+W_k)}{F_{13}(z)F_{13}(w)}\biggr)\biggr] \\
+ \sum_{4\leq k<l\leq N} \frac{F_{12}(z)F_{12}(w)}{F_{13}(z)F_{13}(w)F_{23}(z)F_{23}(w)}\ [F_{3k}(z)F_{3k}(w)]\, [F_{3l}(z)F_{3l}(w)]\  \bigl(Z_k-Z_l\bigr) \bigl(W_k-W_l\bigr)\,.
\end{multline}
This way, one can evaluate the scalar product $\la \{j_i,Z_k\}|\{j_i,W_k\}\ra$ by taking successive derivatives of any of the generating functions \eqref{coh}, \eqref{exp}, \eqref{alg} with respect to $F_{12}, F_{i3}, i=1,2,4,\dotsc,N$, so as to fix the $N$ spins $j_i$.

\section{Generating Functions on The 2-Vertex Graph} \label{sec:GF-calculations}

The series defining the generating function $\cW^{f_J}$ comes from the integration over $\SU(2)$ group elements hidden in the scalar product between coherent intertwiners:
\beq
\cW^{f_J}(w_i,z_i)
&=&
\sum_J
\f{f_J}{J!(J+1)!}\la J,\{\varsigma z_i\}|J,\{w_i\}\ra
\,=\,
\sum_J f_J\sum_{\sum j_i=J}
\f1{\prod_i {(2j_i)!}}\,\la \{j_i,\varsigma z_i\}|\{j_i,w_i\}\ra\nn\\
&=&
\sum_J f_J\sum_{\sum j_i=J}
\f1{\prod_i {(2j_i)!}}\,
\int_{\SU(2)} dg\, [z_i|g|w_i\ra^{2j_i}
\,=\,
\int dg\,\sum_J \f{f_J}{(2J)!}\,(\tr \,Xg)^{2J}\,.
\label{integral}
\eeq
As it was shown in \cite{un2}, one can compute this integral over $\SU(2)$ at fixed $J$ and obtain\footnote{
The integral vanishes when $2J$ is odd and only has a non-zero value for integer values of the total area $J=\sum_i j_i$.
}
\be
\int dg\,(\tr \,Xg)^{2J}
\,=\,
\f1{J!(J+1)!}\,(\det \,X)^J,
\ee
thus reproducing the previous formula for the generating functions.

In the following, we will show that we can evaluate such integrals over $\SU(2)$ as Gaussian integrals and recover the closed formulas for the generating functions, \eqref{coh} and \eqref{alg}. We will use the parametrization \eqref{su2_spinor} of $\SU(2)$ group elements in terms of spinors $g=\f{|z\ra[w|-|z]\la w|}{\sqrt{\la z|z\ra\la w|w\ra}}\,\in\SU(2)$ for $z,w\in\C^2$. Then as shown in \cite{spinor_johannes}, one can reformulate all integrals over $\SU(2)$ as Gaussian integrals over the spinor variables $z$ and $w$.

\subsection{The Schwinger Generating Function}
\label{generating_section}


An arbitrary group element can always be thought as mapping the origin spinor $|\Omega\ra=\left(\begin{smallmatrix} 1 \\0\end{smallmatrix}\right)$ to the arbitrary normalized spinor $|Z\ra$ with $\la Z|Z\ra$,
\be
g\,|\Omega\ra =|Z\ra,\quad
g=|Z\ra\la\Omega|+|Z][\Omega|
=\mat{cc}{Z_0 & -\bZ_1 \\ Z_1 & \bZ_0},
\qquad
\textrm{with}\quad
\la Z|Z\ra = |Z_0|^2+|Z_1|^2=1\,.
\ee
This identifies $\SU(2)$ with the 3-sphere $\cS^3$, and the Haar measure is simply induced by the Lebesgue measure on $\C^2\sim\R^4$,
\be
\int_{\SU(2)} dg \,f(g)
\,=\,
\f1{\pi^2}\int_{\C^2} d^4Z\,\delta(\la Z|Z\ra -1) \,f(Z)
\,=\,
\f1{2\pi^2}\int_{\cS^3} d^3Z\, \,f(Z)\,.
\ee
Now, following \cite{spinor_johannes}, we un-freeze the norm of the spinor $Z$ with a Gaussian weight, to get the following
\begin{res}
\label{gaussian_result}
The integral of a homogeneous polynomial $P(g)$ in $g\in\SU(2)$ of even\footnote{
Integrals over $\SU(2)$ of homogeneous polynomials in $g$ of odd degree  trivially vanish.
} degree  $2J$ can be expressed as a Gaussian integral over $\C^2$,
\be
\int_{\SU(2)} dg \,P(g)
\,=\,
\f1{(J+1)!}\,\int_{\C^2}\f{d^4z}{\pi^2}\,e^{-\la z|z\ra}\,P(z)\,.
\ee
\end{res}

\begin{proof}
Considering an arbitrary function $\psi(\lambda)$ on $\R_+$ normalized to $\int_0^{+\infty}\psi =1$, we can write
\be
\int_{\SU(2)} dg\ P(g)
\,=\,
\f1{2\pi^2}\int_{\cS^3} d^3Z\ P(Z)
\,=\,
\f1{2\pi^2}\int_0^{+\infty}\int_{\cS^3} d\lambda\, d^3Z\ \psi(\lambda)\,P(Z)
\,=\,
\f1{2\pi^2}\int_{\C^2} d^4z\ \f{\psi(\lambda)}{\lambda^{2J+3}}\,P(z),
\ee
where we have made the change of variable $z=\lambda\, Z$ with $\lambda^2=\la z|z\ra$ and taken into account that the polynomial $P$ is homogeneous of degree $2J$. Now, we take $\psi\propto\lambda^{2J+3}e^{-\lambda^2}$ to conclude and show the previous result. The normalization accounts for the $1/(J+1)!$ factor: $\f{2}{(J+1)!}\,\int_0^{+\infty}d\lambda \, \lambda^{2J+3}e^{-\lambda^2}=1$.
\end{proof}

Now, instead of considering a single spinor $Z$ and parameterize group elements as $g=|Z\ra\la\Omega|+|Z][\Omega|$, we would like to use the parametrization $g={|Z\ra\la W|+|Z][ W|}$ with two normalized spinor variables. We can un-freeze the norm of these spinors and it will introduce twice the factor $(J+1)!$. One can show this by either adapting the previous proof or by using the property of the Haar measure to write $dg =dG\,d\tG^{-1}$ for $G,\tG\in\SU(2)$ and parameterize $G=|Z\ra\la\Omega|+|Z][\Omega|$ and $\tG=|W\ra\la\Omega|+|W][\Omega|$. At the end of the day, we take $P(g)=(\tr\,X g)^{2J}$ and we have shown
\be
\int_{\SU(2)} dg \,(\tr\,X g)^{2J}
\,=\,
\f1{(J+1)!^2}\,\int \f{d^4w}{\pi^2}\f{d^4z}{\pi^2}\,e^{-\la z|z\ra-\la w|w\ra}\,
(\la w|X|z\ra+[ w|X|z])^{2J}\,.
\ee
We insert this expression in the integral formula \eqref{integral} for the algebraic generating function,
\be
\cW^{alg}(w_i,z_i)
\,=\,
\int dg\,\sum_J \f{(J+1)!^2}{(2J)!}\,(\tr \,Xg)^{2J}
\,=\,
\int \f{d^4w}{\pi^2}\f{d^4z}{\pi^2}\,e^{-\la z|z\ra-\la w|w\ra}\,
\sum_J \f{1}{(2J)!}(\la w|X|z\ra+[ w|X|z])^{2J}\,.
\ee
Taking into account that all the integrals vanish for odd $(2J)$, we can re-sum over $J$ to get an exponential, which shows the following
\begin{res}
The algebraic generating function can be written as a Gaussian integral over spinor variables,
\be
\cW^{alg}(w_i,z_i)
\,=\,
\sum_J (J+1)!^2\sum_{\sum j_i=J}
\f1{\prod_i {(2j_i)!}}\,\la \{j_i,\varsigma z_i\}|\{j_i,w_i\}\ra
\,=\,
\int \f{d^4w}{\pi^2}\f{d^4z}{\pi^2}\,e^{-\la z|z\ra-\la w|w\ra}\,
e^{\la w|X|z\ra+[ w|X|z]}\,.
\ee
\end{res}

This feature is generalizable to arbitrary graphs beyond our simple 2-vertex graph. This factor is for instance used in \cite{jeff}  to evaluate exactly the generating function and recover the previous results \cite{schwinger:52, bargmann:62, wu-9j, huang-wu-2j-15j, labarthe, schnetz, garoufalidis, costantino-generating}. The advantage of this reformulation is that we can compute easily the Gaussian integral,
\be
\begin{aligned}
\int \f{d^4w}{\pi^2}\f{d^4z}{\pi^2}\,e^{-\la z|z\ra-\la w|w\ra}\,
e^{\la w|X|z\ra+[ w|X|z]}
&=
\int \f{d^4w}{\pi^2}\f{d^4z}{\pi^2}\,e^{-\la z|z\ra-\la w|w\ra}\,
e^{\la w|X|z\ra+\la z|\eps^{-1}({}^tX)\eps|w\ra}\\
&=
\f1{\det M}
\qquad\textrm{with}\quad
M=\mat{c|c}{\id & -X \\ \hline -\eps^{-1}({}^tX)\eps & \id}\,,
\end{aligned}
\label{alg_gaussian}
\ee
Since the identity $X\eps^{-1}({}^tX)\eps=\,(\det X)\,\id$ holds for arbitrary $2\times 2$ matrices, one computes the determinant of the matrix by block:
\be\nn
\det M
\,=\,
\det_{2\times 2}(\id-X\eps^{-1}({}^tX)\eps)
\,=\,
\det_{2\times 2}((1-\det X))\id)
\,=\,
(1-\det X)^2.
\ee
We have thus reproduced the expected result for the algebraic generating function, $\cW^{alg}(w_i,z_i)= (1-\det X)^{-2}$.

As described in \cite{jeff}, these Gaussian integral techniques can be generalized to arbitrary graphs to evaluate the algebraic generating function explicitly in terms of  a block determinant. Expanding this determinant as a sum over cycles, one then recovers a closed formula for this generating function as a rational function in the spinor variables as mentioned in \cite{recursion_spinor} and shown in \cite{schwinger:52, bargmann:62,wu-9j, huang-wu-2j-15j, labarthe, schnetz, garoufalidis, costantino-generating}.

\subsection{Coherent Spin Network and the Geometric Generating Function}
\label{compute}


Considering the coherent spin network evaluation, we plug $f_J=1$ in the generic integral formulation \eqref{integral} and write
\be
\cW(w_i,z_i)
\,=\,
\int dg\,\sum_J \f1{(2J)!}\,(\tr X g)^{2J}
\,=\,
\int dg\,e^{\tr X g}\,.
\ee
There are various ways to compute this integral. We can either write explicitly $g$ in terms of its complex matrix elements or parameterize it in terms of spinor variables. As we show below, both cases lead to Gaussian integrals, but with a unit norm constraint. As explained in appendix \ref{gaussian}, while straightforward Gaussian integrals give rational function in the parameters, such constrained Gaussian integrals typically lead to Bessel functions.

\paragraph{First method.} It has been reported in the appendix of \cite{un2}, by computing $\int dg\,(\tr X g)^{2J}$ directly on the 3-sphere.

\paragraph{Second method.} Let us start with writing the group element $g$ in terms of its matrix elements:
\be
g=\begin{pmatrix} Z_0 & -\bZ_1 \\ Z_1 & \bZ_0 \end{pmatrix},
\qquad
\text{with}\quad
\det\,g=|Z_0|^2+|Z_1|^2=1,
\qquad
dg = \,\f1{\pi^2}\delta(|Z_0|^2+|Z_1|^2-1)\,d^2Z_0d^2Z_1\,.
\ee
Due to the spherical constraint in the measure, one does not change integrals with respect to $dg$ when an extra factor $e^{-a(|Z_0|^2+|Z_1|^2-1)}$ is introduced into the integrand. Then we Fourier transform the spherical constraint to get
\be
\begin{aligned}
\int dg\,e^{\tr X g}
&= \f1{\pi^2} \int d^2Z_0d^2Z_1\ e^{\tr X g}\,\int_\R \frac{dT}{2\pi}\,e^{-(a+iT)(|Z_0|^2+|Z_1|^2)}\,e^{a+iT}\\
&= \f1{\pi^2} \int d^2Z_0d^2Z_1\ e^{\tr X g}\,\int_{a-i\infty}^{a+i\infty} \frac{ds}{2\pi}\,e^{-s(|Z_0|^2+|Z_1|^2)}\,e^{s}
\end{aligned}
\ee
If we take $a=\Re(s)>0$, one can safely exchange the integral over $s$ with the one over $\C^2$. Also writing $X= \left(\begin{smallmatrix} \alpha &\beta\\ \gamma &\delta\end{smallmatrix}\right)$ and expanding in terms of the real and imaginary parts of $Z_0$ and $Z_1$, we obtain a well-defined Gaussian integral which is easily performed,
\be
\int dg\,e^{\tr X g} = \int_{a-i\infty}^{a+i\infty} \frac{ds}{2\pi}\,e^s\,\frac{e^{\frac{\det X}{s}}}{s^2}.
\ee
This is an evaluation of the inverse Laplace transform of $e^{\det X/s}/s^2$. We expand $e^{\det X/s}$ into powers of $\frac{\det X}{s}$ and use
\be
\int_{a-i\infty}^{a+i\infty} \frac{ds}{2\pi}\,e^s\,\frac{1}{s^{n+2}} = \frac1{(n+1)!},
\ee
which can be found by the residue theorem typically, considering the pole at $T=ia$ for $s=a+iT$ and $a>0$. This way one obtains
\be
\int dg\,e^{\tr X g} = \sum_{n\geq0} \frac{(\det X)^n}{n!} \int_{a-i\infty}^{a+i\infty} \frac{ds}{2\pi}\,e^s\,\frac{1}{s^{n+2}} = \sum_{n\geq0} \frac{(\det X)^n}{n!(n+1)!},
\ee
which is indeed a series expansion of $I_1(2\sqrt{\det X})/\sqrt{\det X}$.

\paragraph{Third method.} We can also use a technique similar to the one used previously to compute the algebraic generating function, by mapping the integral over $g\in\SU(2)$ to a Gaussian integral over the spinor variables $z,w\in\C^4$. Indeed,
\beq
\int dg\,e^{\tr\,X g}
&=&
\int_{\la w|w\ra=\la z|z\ra=1} \f{d^3w}{\pi^2}\f{d^3z}{\pi^2}\,e^{\la w|X|z\ra+[ w|X|z]} \nn\\
&=&
\int \f{e^{-iT}dT}{2\pi} \f{e^{-i\tT}d\tT}{2\pi}
\int \f{d^4w}{\pi^2}\f{d^4z}{\pi^2}\ e^{iT\la z|z\ra+i\tT\la w|w\ra}\,e^{\la w|X|z\ra+\la z|\eps^{-1}\,{}^tX\eps|w\ra}\,.
\eeq
Hence we again face a Gaussian integral, very similar to the one found in the case of the Schwinger's generating function \eqref{alg_gaussian}. However this one should be regularized as in the second method to make sure it is convergent. Instead of reproducing the full calculation, let us simply perform the Gaussian integral formally,
\be
\int dg\,e^{\tr X g}
\,=\,
\int \f{e^{-iT}dT}{2\pi}\, \f{e^{-i\tT}d\tT}{2\pi}\ \f1{(T\tT+\det X)^2}\,.
\ee

\paragraph{Relation between the Schwinger and the geometric generating functions.} What those representations of $\cW(z_i,w_i)$ teach us is that we can express the coherent spin network evaluation as a Fourier transform\footnote{Due to the singularity of the integrand after the Gaussian integral, the Fourier transform is ambiguous. The correct way of understanding it is \emph{not} as the principal value of the integral. As we have seen in the second method, it is necessary to introduce an extra factor into the integrand to make the Gaussian integral well-defined. Then this induces a shift of the poles in the complex plane. These integrals should thus be seen as inverse Laplace transforms rather than Fourier transforms.} of the Schwinger's generating function appropriately rescaled. In order to see this, let us remember that $\det X=\sum_{i< j} F_{ij}\tF_{ij}$ where $F_{ij}\equiv F_{ij}(z)$ and $\tF_{ij}\equiv F_{ij}(w)$. We can then consider both generating functions as functions in $F_{ij}$ and $\tF_{ij}$. Carefully considering the Gaussian integral above, we have:
\be
\cW(F_{ij},\tF_{ij})
\,=\,
\int \f{e^{-iT}dT}{2\pi} \f{e^{-i\tT}d\tT}{2\pi}\,
\f1{(iTi\tT)^2}\,\cW^{\text{alg}}\Bigl(\f{F_{ij}}{iT},\f{\tF_{ij}}{i\tT}\Bigr)\,.
\ee
We discuss how to generalize this feature to arbitrary graph in section \ref{future}, allowing to deduce the more complicated evaluation of the coherent spin network from the algebraic generating function. This procedure relies on the reformulation of the coherent spin  network evaluation as integrals over $\SU(2)$, which can be written as integrals over normalized spinors and then as Gaussian integrals over these spinor variables after Fourier transforming the constraint of unit norm. On the other hand, the algebraic generating function is directly a Gaussian integral over the spinors, it can thus be computed explicitly and  can then be used as a first step in order to compute the coherent spin network evaluation.

\subsection{Towards spin networks with non-trivial holonomies} \label{sec:add-holonomy}

The results presented so far focus on the evaluation of the spin network function with coherent intertwiners and trivial holonomies. In this section, we suggest a generalization to non-trivial holonomies. From the definition \eqref{modified def} and using the translation invariance of the Haar measure,
\be
\vphi_{w_i,z_i}^{f_J}(g_i)
\,=\,
\sum_{\{j_i\}}
\f{f_{\sum_i j_i}}{\prod_i {(2j_i)!}}\,\int dg \,dh\ \prod_{i=1}^N [j_i,z_i| g\,h g_i h^{-1}\,|j_i,w_i\rangle\,.
\ee
This is equivalent to evaluating the spin network function on trivial holonomies but with rotated spinors $\tl{w_i} = h g_i h^{-1} w_i$, and averaging over the adjoint action of $h$,
\be
\vphi_{w_i,z_i}^{f_J}(k)
\,=\,
\int dh\ \cW^{f_J}(h g_i h^{-1}w_i, z_i)\,.
\ee

This simplifies when all links of the 2-point graph but one have the same holonomy. Using $\SU(2)$ invariance, one can take $g_1 = k \in\SU(2)$ and $g_i=\mathbbm{I}$ for $i=2,\dotsc,N$. Then
\be
\vphi_{w_i,z_i}^{f_J}(k)
\,=\,
\sum_{\{j_i\}}
\f{f_{\sum_i j_i}}{\prod_i {(2j_i)!}}\,\int dg \,dh\ [j_1,z_1| g\,hkh^{-1}\,|j_1,w_1\rangle \prod_{i=2}^N [j_i, z_i|g|j_i,w_i\ra\,.
\ee
The integral of the adjoint action on $k$ shows that $\vphi_{w_i,z_i}^{f_J}(k)$ only depends on the conjugation class of the holonomy $k$. Moreover, this averaging can be explicitly performed in the spin expansion using the orthogonality of the matrix elements of $h$,
\be
\int dh\ [j_1,z_1| g\,hkh^{-1}\,|j_1,w_1\rangle = \frac{\chi_{j_1}(k)}{2j_1+1}\ [j_1,z_1| g |j_1,w_1\rangle\;,
\ee
where $\chi_j$ is the character in the representation of spin $j$, which reads $\chi_j(k) = \frac{\sin (2j+1)\theta_k}{\sin \theta_k}$ if $\theta_k$ is the class angle of $k$. In the same way the spin dependence of the matrix element $[j_1,z_1| g |j_1,w_1\rangle = [z_1| g |w_1\rangle^{2j_1}$ is only via the exponent, one tries to rewrite the full $j_1$ dependence of the above formula with an exponential. For instance
\be
\frac{\chi_{j_1}(k)}{2j_1+1} = \frac{1}{2i\,\sin \theta_k}\int_{\R_+} dt\ e^{-t+i\theta_k} \Bigl(e^{-t+i\theta_k}\Bigr)^{2j_1} - e^{-t-i\theta_k} \Bigl(e^{-t-i\theta_k}\Bigr)^{2j_1};,
\ee
Hence
\begin{align}
& \begin{aligned} \vphi_{w_i,z_i}^{f_J}(k) =  \int_{\R_+} \frac{dt\,e^{-t}}{2i\,\sin \theta_k} \sum_{\{j_i\}}
\f{f_{\sum_i j_i}}{\prod_i {(2j_i)!}}\, \int dg\
 &\prod_{i=2}^N [j_i, z_i|g|j_i,w_i\ra \\
&\times \left(e^{i\theta_k} [j_1,z_1| g |j_1, e^{-t+i\theta_k}w_1\rangle + e^{-i\theta_k} [j_1,z_1| g |j_1, e^{-t-i\theta_k}w_1\rangle\right) \,, \end{aligned}
\\
&\phantom{\vphi_{w_i,z_i}^{f_J}(k)} = \frac{1}{2i\,\sin \theta_k} \int_{\R_+} dt\,e^{-t}\,\Bigl[e^{i\theta_k}\ \cW^{f_J}(e^{-t+i\theta_k}w_1,z_1, w_2,z_2,\dotsc) - e^{-i\theta_k}\ \cW^{f_J}(e^{-t-i\theta_k}w_1,z_1, w_2,z_2,\dotsc)\Bigr]\;.
\end{align}
This means that the curvature induced by a single non-trivial holonomy on one link can be accounted for as a scalar multiplication of the spinor $w_1$ (or equivalently $z_1$) by $e^{-t\pm i\theta_k}$, and an integral over the `Schwinger' parameter $t$.




\section{Stationary Point Analysis and Geometric Interpretation} \label{sec:saddlepoint}



\setcounter{paragraph}{0}

Let us consider the series defining the coherent spin network evaluation and look at the probability distribution it induces on the total area $J$ and the individual spins $j_i$.

\paragraph{Probability distribution on the total area $J$.} We start with the series $\cW(z_i,w_i)=\sum_J \f{(\det X)^J}{J!(J+1)!}$. The behavior of these terms at large total area $J$ is found using the Stirling formula,
\be
\f{(\det X)^J}{J!(J+1)!}
\,\underset{J\gg 1}{\sim}\,
\f{1}{2\pi J^2}\,e^{\phi(J)}\,,
\qquad
\phi(J)\equiv J\ln\det X -2(J\ln J-J)\,.
\ee
This probability distribution on $J$ defined by the expansion of the Bessel function is very similar to a Poisson distribution. This distribution is peaked in $J$ on the stationary point
satisfying $\pp_J \phi=0$, that is:
\be
J^2=\det X\,.
\ee
Thus as soon as $\det X$ is large, the Stirling approximation is valid and we obtain a Gaussian-like distribution around this stationary point $J\sim\sqrt{\det X}$. Let us point out that $\det X$ is in practice complex and this should be considered more exactly as a saddle point. Computing the second derivative $\pp_J^2\phi=-2/J$ and the corresponding saddle point approximation leads us back to the standard asymptotic approximation for the Bessel function:
\be
\cW(w_i,z_i)\sim \f{1}{\sqrt{4\pi}(\sqrt{\det X})^{\f32}}\,e^{2\sqrt{\det X}},.
\ee
A special case is for flat configurations of the classical labels, i.e. when all the spinors $z_i$ are related to the spinors $w_i$ by a single $\SU(2)$ transformation for all edges, $|z_i]=g\,|w_i\ra$ for all $i$'s for a given group element $g\in\SU(2)$. In this case, the determinant of $X$ is strictly positive and equal to the square of the total area:
\be
|z_i]=g\,|w_i\ra
\quad
\Rightarrow\quad
X= g^{-1}\,\sum_i|z_i][z_i|=g^{-1}\, A(z_i)\id
,\quad
\det X=\det \sum_i|z_i][z_i| = A(z_i)^2=A(w_i)^2\,.
\ee
Then the stationary point dominating our series for the coherent spin network evaluation is given simply by the sum $J$ of the spins equal to the total classical area, $J=A(z_i)$  as expected.

\paragraph{Probability distribution on the spins.} We can go further and study the finer structure of the coherent intertwiners and spin networks. We would like indeed to describe the probability distribution for the individual spins $j_i$ living on each edge of the graph. Since the generating function can be decomposed in terms of LS intertwiners as
\be
\cW(w_i,z_i)=\sum_{j_i}\f{1}{\prod_i (2j_i)!}\la\{j_i,\varsigma z_i\} |\{j_i,w_i\}\ra,
\label{W2}
\ee
the spins $j_i$ follow approximatively Poisson distributions, as explained in \cite{un4}, if we neglect the group averaging and simply assume that the scalar product $\la\{j_i,\varsigma z_i\} |\{j_i,w_i\}\ra$ goes as $\prod_i [z_i|w_i\ra^{2j_i}$. This is a crude approximation, but it  represents rather well what actually happens. Indeed, here we can compute exactly the evaluation $\cW(w_i,z_i)$ and use it to extract exact probability distribution.

Starting from the expression of $\det X$ in terms of the spinors $z_i$ and $w_i$, we write explicitly the series
\be
\cW(w_i,z_i)
\,=\,
\sum_J \f1{J!(J+1)!}\biggl(\sum_{i<j}F_{ij}(z)F_{ij}(w)\biggr)^J
\,=\,
\sum_{k_{ij}}
\f{1}{(\sum_{i<j}k_{ij}\,+1)!}
\f1{\prod_{i<j} k_{ij}!}
\prod_{i<j}F_{ij}(z)^{k_{ij}}F_{ij}(w)^{k_{ij}}
\,.
\ee
The observables $F_{ij}(z)$ and $F_{ij}(w)$ are holomorphic respectively in $z_i,z_j$ and in $w_i,w_j$. Considering the definition \eqref{W2}, the terms of the series corresponding to fixed spins $j_i$ are homogeneous of degree $2j_i$ in $z_i$ and $w_i$. This is therefore easy to identify the terms corresponding to fixed $j_i$ and regroup them according to
\be
2j_i=\sum_{j\ne i} k_{ij},
\qquad\textrm{with}
\quad
k_{ji}=k_{ij}\,.
\ee
This method was actually used in \cite{un2} to compute the scalar product $\la\{j_i,z_i\} |\{j_i,w_i\}\ra$ between LS intertwiners.
The spins $j_i$ obviously satisfy $\sum_i 2j_i=\sum_{i\ne j}k_{ij}=2J$. However, we have $N$ spin labels $j_i$ compared to the $N(N-1)/2$ integers $k_{i<j}$. Hence the extra integers that we are summing over should correspond to internal degrees of freedom of the intertwiners (and maybe they can be used to define a new basis of intertwiners).

From the present perspective, the $k$'s appear as much more natural variables than the spins $j_i$. We thus propose to study the probability distribution of the $k$'s and deduce from it the behavior of the spins. As before, we use Stirling approximation for the factorials (where we discard the $\sqrt{2\pi k_{ij}}$ factors which are irrelevant for the present discussion):
\be
\cW(w_i,z_i)
\,\sim\,
\sum_{k_{ij}}
\f1{\sum_{i<j}k_{ij}}\,
e^{\sum_{i<j}k_{ij}\ln(F_{ij}(z)F_{ij}(w))-k_{ij}(\ln k_{ij} -1)\,-(\sum_{i<j}k_{ij})\left[\ln (\sum_{i<j}k_{ij}) -1\right]}\,.
\ee
Looking for the stationary points of the exponent gives the equations
\be
\forall i,j,\quad \ln k_{ij}J \,=\, \ln F_{ij}(z)F_{ij}(w)\,,
\ee
with $J=\sum_{i<j}k_{ij}$. Discarding the log and summing over $i<j$ gives:
$$
J^2=\sum_{i<j} F_{ij}(z)F_{ij}(w)=\det X,
$$
as before. Thus we have a unique stationary point for each set of classical spinors $z_i,w_i$ and it is  given by:
\be
k_{ij}= \f{F_{ij}(z)F_{ij}(w)}{\sqrt{\sum_{k<l} F_{kl}(z)F_{kl}(w)}}\,.
\ee

It is easier to understand the geometrical meaning of this fixed point in the flat case, when $|z_i]=g|w_i\ra$ or equivalently $F_{ij}(z)=\bF_{ij}(w)$. In that case, converting the previous formula in terms of the 3-vectors $\vV_i=\vV(z_i)$ and remembering the closure constraint $\sum_i \vV_i=0$, we get the much simplified following expression:
\be
k_{ij}= \f12\f{|\vV_i||\vV_j|-\vV_i\cdot\vV_j}{A(z)},\qquad
A(z)=\f12\sum_i |\vV_i|\,.
\ee
Translating this in terms of the spins $j$, we derive:
\be
2j_i=\sum_{j} k_{ij}=|\vV_i|,
\ee
which is the expected classical values for the spins.
Then the coherent spin network evaluation defines approximate Gaussian distributions peaked on these classical values. Moreover, as we have already seen in the case of the series in $J$, computing the Hessian and saddle point approximation for this series will provide us with a good asymptotic approximation for the exact Bessel function expression of $\cW(w_i,z_i)$.

\paragraph{Saddle point evaluation of the Schwinger's generating function.} We can perform the same analysis on the algebraic generating function,
\beq
\cW^{alg}(w_i,z_i)
&=&
\sum_J (J+1)!
\sum_{\sum_{i<j}k_{ij}=J}
\f1{\prod_{i<j} k_{ij}!}
\prod_{i<j}F_{ij}(z)^{k_{ij}}F_{ij}(w)^{k_{ij}}\nn\\
&\sim&
\sum_{k_{ij}}
\f1{\sum_{i<j}k_{ij}}\,
e^{\sum_{i<j}k_{ij}\ln(F_{ij}(z)F_{ij}(w))-k_{ij}(\ln k_{ij} -1)\,+(\sum_{i<j}k_{ij})\left[\ln (\sum_{i<j}k_{ij}) -1\right]}\,.
\eeq
All that changes is the $(J+1)!$ factors. However the scaling of the stationary point crucially depends on this factor:
\be
\forall i,j,\quad \ln \f{k_{ij}}{J} \,=\, \ln F_{ij}(z)F_{ij}(w),
\qquad
J=\sum_{i<j}k_{ij}\,.
\ee
This means that if there is a stationary point, then there is actually a stationary line, since the space of solutions is invariant under global rescaling of all the $k_{ij}$ by an arbitrary constant. Nevertheless, we can see by summing over $i<j$ that the existence of this stationary line requires
\be
\sum_{i<j}\f{k_{ij}}{J}=1=\sum_{i<j}F_{ij}(z)F_{ij}(w).
\ee
This implies that, when $\det X\ne 1$, we do not have any contribution and the saddle point approximation fails. On the other hand, when $\det X=1$, we get a fixed line invariant under rescaling, which implies a divergence. This stationary line and its associated divergence are easily interpreted from the exact form of the generating function $1/(1-\det X)^{2}$, as due to the pole on $\det X=1$.

\paragraph{Saddle point evaluation of the exponential generating function.} We can perform similarly the same analysis on the exponential generating function $\cW^{exp}(w_i,z_i)$.
This time, the factors $(J+1)!$ simply disappear. And the stationary point equation simplifies to:
\be
\forall i,j,\, k_{ij}=F_{ij}(z)F_{ij}(w)\,.
\ee
Thus the stationary point always exists unlike for Schwinger's generating function. But it differs from the case of the coherent spin network evaluation $\cW(w_i,z_i)$ by an overall area factor. Indeed considering the special flat case as before, when $|z_i]=g|w_i\ra$ with the same group element $g$ for all $i$'s, we get:
\be
k_{ij}= \f12\left(|\vV_i||\vV_j|-\vV_i\cdot\vV_j\right),\qquad
2j_i=\sum_{j} k_{ij}=|\vV_i|A(z),\qquad
J=\sum_i j_i = A(z)^2\,.
\ee
Thus the spins $j_i$ are peaked on a classical geometry rescaled  by a total area $A(z)$ factor. It is unclear what use could such a generating function have.

\paragraph{Comparison of the different generating functions.} This analysis shows that the coherent spin network evaluation has a nice geometrical interpretation in terms of its series being peaked on the correct classical discrete geometry and this is why we call it the \emph{geometric generating function}, while the Schwinger's choice for the generating function does not admit such a natural geometrical interpretation. The exponential generating function is a bit similar to the geometric one, but the saddle point contribution exhibits a surprising rescaling by the total area.

\section{Wheeler-DeWitt equations for the flat dynamics} \label{sec:pde}

Spinors parametrize our phase space. Therefore they can be used to build coherent states and also observables which become operators upon quantization. In the coherent basis, these operators translate to differential operators acting on the spinor labels of the wave-function \cite{sfcosmo}
\be
\cO \psi_z(g) = \cD_z \psi_z(g)\,.
\ee

The generating functions we have considered in the previous sections are evaluations of wave-functions on the identity $\psi_z(\mathbbm{I})$. Physically, this means restricting to flat space, where Wilson loops are trivial. Hence, one should characterize these states by an equation which would then be a Wheeler-DeWitt equation for a Hamiltonian constraint corresponding to flat space.

To this aim, we use the construction of the Hamiltonian constraint for the topologically flat model which was introduced  in \cite{recursion3d, recursion4d} in the loop quantum gravity context and generalized to spinors in \cite{recursion_spinor}. The idea is that given two spinors defined at the same node their product is usually not invariant under the rotation of one of them by a Wilson loop. However, when Wilson loops are trivial one can take as a constraint the invariance of their product. For the graph with two vertices, $[w_i| w_j\rangle = [w_i| g_i^{-1} g_j |w_j\rangle$ as $g_i^{-1} g_j$ is the holonomy around the closed path along the lines $i$ and $j$. By definition, on our phase space parametrized by spinors, holonomies are functions of spinors, \eqref{su2_spinor}, hence we propose \eqref{flat-constraint} as a constraint,
\be \nn
F_{ij}(w) - \bF_{ij}(z) = 0\;.
\ee

Now we look at the quantization of this constraint. First, we check that its annihilates the evaluations $\cW^{f_J}(w_i,z_i)$ and then we write the differential equation it generates explicitly on the evaluation with the choice $f_J=1$.

The first step is almost trivial. One notices that $\hFd_{ij}(z)$ acts on the \emph{dualized} intertwiner $\langle \{\varsigma z_k\}|$,
\be
\hFd_{ij}(z)\ \cW^{f_J}(w_k,z_k) = \sum_J \frac{f_J}{J! (J+1)!} \left(\langle J, \{\varsigma z_k\}| \hF_{ij}\right)\,|J,\{w_k\}\rangle\;.
\ee
This is obviously the same as $\hF_{ij}(w) \cW(w_k,z_k)$. In the following we derive the differential equation for $f_J=1$.

We have already found in equation \eqref{hF_CI} that $\hF_{ij}(w)$ is diagonal on coherent intertwiners, $\hF_{ij}(w)\,|\{w_k\}\ra= [w_i|w_j\ra\,|\{w_k\}\ra = F_{ij}(w)\,|\{w_k\}\ra$. Similarly one evaluates
\be
\langle \{\varsigma z_k\}|\,\hF_{ij}(z) = \left[ \partial_{z_i}|\partial_{z_j}\right\rangle\ \langle \{\varsigma z_k\}|\;,
\ee
hence
\be \label{wdw}
\Bigl(\left[ \partial_{z_i}|\partial_{z_j}\right\rangle - [w_i|w_j\rangle\Bigr)\ \cW(w_k,z_k) =0\;.
\ee
This equation is remarkable simple. The constraint also implies equations on $\cW^{f_J}$ with other choices of $f_J$ but there are more complicated. Indeed, one has then to generate the weight function $f_J = (J+1)!^\alpha$ and this can be done by additional derivatives with respect to the spinors. In essence, the equation is similar but it receives higher order derivatives to account for the weight function. This was done explicitly in \cite{recursion_spinor} for the generating function of 6j-symbols with $f_J=(J+1)!$.

Let us now forget that we know how to calculate $\cW(w_i,z_i)$ in a closed form and instead make some ansatz to reduce the above equation,
\be \label{ansatz detX}
\cW(w_i,z_i) = \phi (\det X)\;,
\ee
where we remind the reader that $X = \sum_i |w_i\rangle [z_i|$ is a $2\times2$ matrix. This is actually a crucial assumption, which turns out to hold for any generating function $\cW^{f_J}$. The fundamental reason is that the scalar product of coherent intertwiners at fixed total area is a function of $\det X$, $\langle J,\{\varsigma z_i\}| J,\{w_i\}\rangle = (\det X)^J$.
%
%
To prove this ansatz \eqref{ansatz detX} without calculating this scalar product, we haven't found any direct proof from the differential equations \eqref{wdw}. On the other hand, we show below that $\cW$ satisfies further differential equations, which reflect its invariance under $\U(N)$ transformations, from which it is straightforward to prove that $\cW$ must be a function of solely $\det X$. The equivalence between the equations \eqref{wdw} and the $\U(N)$ equations  seems to be true under the assumption of an extra ``closure constraint" differential equation but we have not been able to prove it explicitly, as we expand upon below.

The ansatz \eqref{ansatz detX} reduces \eqref{wdw} to an ordinary differential equation on $\phi$,
\be
x\,\phi'' + 2\,\phi' -\phi=0\;.
\ee
We have used $[\partial_{z_2}\det X|\partial_{z_1} \det X\rangle = F_{21}(w)\,\det X$, and $[\partial_{z_2}|\partial_{z_1}\rangle (\det X)^J = J(J+1) F_{21}(w)\,(\det X)^{J-1}$. Introducing the change of function $I(y) = \frac{y}{2}\,\phi(\frac{y^2}{4})$ leads to a well-known equation on $I$,
\be
y^2\,I'' + y\,I' - (y^2 +1)\,I = 0\;,
\ee
i.e. the modified Bessel's equation of order 1. We conclude that $I=I_1$ and thus $\cW = I_1(2\sqrt{\det X})/\sqrt{\det X}$.

\smallskip

Let us come back to the ansatz \eqref{ansatz detX} and to proving that $\cW$ is a function of $\det X$. To make things more explicit and easier to handle, we drop the notation $|\ra$ and $|]$ for the spinors and use explicitly the spinor indices $A=0,1$. The differential equations \eqref{wdw} then read:
\be
\epsilon_{AB}\pp_{z_i^A}\pp_{z_j^B}\cW= \eps_{AB}w_i^Aw_j^B\cW,
\qquad
\epsilon_{AB}\pp_{w_i^A}\pp_{w_j^B}\cW= \eps_{AB}z_i^Az_j^B\cW,
\ee
where we are including the reverse equations where we have swapped the role of the $z$'s and $w$'s. Following the ideas of \cite{2vertex,spinor}, we identify further differential equations satisfied by the the coherent spin network evaluation:
\be\nn
\forall i,j,\quad
\sum_A z_i^A\pp_{z_j^A}\cW
\,=\,
\sum_A w_j^A\pp_{w_i^A}\cW\,.
\ee
Using the expression of the generating function as an integral over $\SU(2)$, explicitly $\cW(w_k,z_k)=\int dg\,\exp\,\sum_k [z_k|g|w_k\ra$, it is actually almost trivial to check that it satisfies the differential equations \eqref{wdw} and these new equations:
\begin{gather}\nn
\left[ \partial_{z_i}|\partial_{z_j}\right\rangle\,\cW
\,=\,
\int dg\,-[w_j|g^{-1}g|w_i\ra\,e^{\sum_k [z_k|g|w_k\ra}
\,=\,
[w_i|w_j\rangle\,\cW\,,\\
\sum_A z_i^A\pp_{z_j^A}\,\cW
\,=\,
\int dg\,[z_i|g|w_j\ra\,e^{\sum_k [z_k|g|w_k\ra}
\,=\,
\sum_A w_j^A\pp_{w_i^A}\,\cW\,.\nn
\end{gather}
The interest of the differential operators $\cE_{ij}\equiv \sum_A z_i^A\pp_{z_j^A}- w_j^A\pp_{w_i^A}$ is that they form a closed $\u(N)$ algebra. As exploited in \cite{2vertex,spinor}, they generate the following $\U(N)$ action on the spinor variables:
\be
z_i,\,w_i
\quad\mapsto \quad
(Uz)_i=\sum_j U_{ij}z_j,\,
(\bar{U}w)_i=\sum_j \bar{U}_{ij}w_j,
\qquad\textrm{for}\quad
U\in\U(N).
\ee
There are only 3 (quadratic) $\U(N)$ invariants (which must also be invariant under $\SU(2)$) from which one can generate all $\U(N)$ invariants:
\be
\sum_{ij} \la z_i|z_j \ra\la w_i|w_j \ra,\quad
\sum_{ij} [ z_i|z_j \ra [ w_i|w_j \ra,\quad
\sum_{ij} \la z_i|z_j ]\la w_i|w_j ]\,.
\ee
This is rather natural and can be proved decomposing the $\U(N)$ action into irreducible representations \cite{2vertex,spinor}. Since $\cW(w_i,z_i)$ is holomorphic in both $z_i$ and $w_i$, it must be a function of the second invariant, recognized as $\det X$.

Moreover, it seems possible to show that  the $F$-equations \eqref{wdw} and the $\U(N)$ differential equations are equivalent under assuming that the generating function $\cW$  satisfies the quantum equivalent of the closure constraint,
\be
\sum_k z_k^B\pp_{z_k^C} \cW = \delta_{BC}\,\f12\sum_A\sum_k z_k^A\pp_{z_k^A} \cW, 
\ee
and similarly on the $w$'s. It is easy to show that $\cW$ satisfies these equations by using its formula as an $\SU(2)$ integral\footnote{
For the closure constraint equation, we write:
\be
\sum_i z_i^B\pp_{z_i^C} \vsigma_{BC}\,\cW
\,=\,
\int dg\,\sum_i [z_i|\vsigma g|w_i\ra\,e^{\sum_k [z_k|g|w_k\ra}
\,=\,0\,,
\ee
since this is a total derivative on $\SU(2)$. Since the 3 components of the operator $\sum_i z_i^B\pp_{z_i^C}$
projected on the Pauli matrices vanish, only its trace survives.
} as before.

Indeed, following the ideas of \cite{spinor}, it seems natural to compose the equations \eqref{wdw} as a $N\times N$ matrix multiplication. Using the closure constraint to simplify this product, we are led to the $\U(N)$ differential equations up to global factors $\sum_k z_k^A\pp_{z_k^A}$ and $\sum_k w_k^A\pp_{w_k^A}$. We do not go into more details since this does not seem to be a crucial point. What is important is that the generating function $\cW$ satisfies both the $F$-equations and the $\U(N)$ equations, and that they form all together a closed algebra (thus not generating further differential equations satisfied by $\cW$).

\smallskip

To conclude this section, we note that simpler equations would hold if one puts aside the Pl\"ucker relations and expresses the generating functions as functions of the variables $F_{ij}$ directly instead of spinors. Then $\det X$ satisfies an obvious differential equation,
\be
\frac{\partial \det X}{\partial F_{ij}(z)} = F_{ij}(w)\;,
\ee
which implies differential equations for arbitrary generating functions on the 2-vertex graph.

\section{Generating Functions of Spin Network Evaluation on Arbitrary Graphs} \label{future}

\subsection{Coherent Spin Networks, Evaluation and Differential Equations}

Up to now, we have discussed in great details the coherent spin network evaluation and generating functions on the 2-vertex graph. In this section, we would like to generalize our framework to arbitrary graphs and present some potential ways to explicitly get the coherent spin network evaluation. Such a computation is essential to spinfoam models since the spinfoam transition amplitudes between quantum states of geometry are defined as some product of coherent spin network evaluation. It is therefore crucial to understand how one could compute them or characterize them through some differential equations.

Considering an arbitrary graph $\Gamma$, we consider a set of spinors $z_e^v$, one for each vertex $v$ and every edge $e$ attached to $v$. That is we have two spinors $z_e^{s(e)}$ and $z_e^{t(e)}$ for each edge $e$, corresponding the source and target vertices $s(e)$ and $t(e)$ of the edge.We now construct the coherent intertwiners $\{\varsigma^{\eps_v(e)} z^v_e\}\ra$ around each vertex $v$.
We recall the weight in the total spin $J$ in their definition:
$$
|\{ z_i\}\ra
\,=\,
\sum_J \f1{\sqrt{J!(J+1)!}}\,|J,\{ z_i\}\ra
\,=\,
\sum_{j_i}\f1{\sqrt{ \prod_i (2j_i)!}}\,|\{ j_i,z_i\}\ra\,.
$$
The power $\eps_v(e)$ is 0 if $v$ is the source vertex of the edge while it is 1 if $v$ is the target vertex, that is we switch the orientation of the spinor if $v=t(e)$. Then we define the coherent spin network function by gluing those coherent intertwiners together. As obtained in \cite{un4}, we get:
\be
\varphi_{\Gamma,z^v_e}(g_e)
=
\int_{\SU(2)^V} [dh_v]\,
e^{\sum_e[z_e|h_{t(e)}^{-1}g_eh_{s(e)}|w_e\ra}\,,
\ee
where we used the notation $w_e\equiv z_e^{s(e)}$ and $z_e\equiv z_e^{t(e)}$ similarly to the case of 2-vertex graph.

The geometric generating function is defined as the evaluation of this coherent spin network wave-function at the identity:
\be
\cW(z^v_e)
\,\equiv\,
\varphi_{\Gamma,z^v_e}(\id)
\,=\,
\int_{\SU(2)^V} [dh_v]\,
e^{\sum_e[z_e|h_{t(e)}^{-1}h_{s(e)}|w_e\ra}\,.
\ee

\begin{figure}[h]
\begin{center}
\includegraphics[height=45mm]{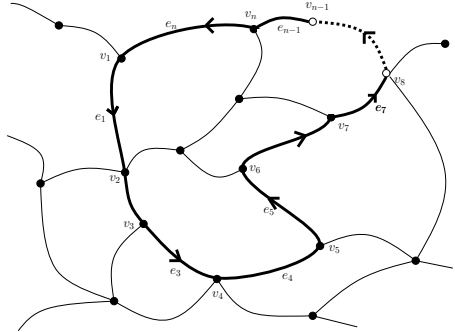}
\caption{A loop $v_1\overset{e_1}{\arr}v_2\dots v_n\overset{e_n}{\arr}v_1$ in the graph $\Gamma$ with spinors $z_i,w_i$ living at each end of every edge. {\it By courtesy of I\~naki Garay from \cite{spinor}.}\label{loop}}
\end{center}
\end{figure}

It is not obvious how to integrate over these $\SU(2)$ group elements living at the vertices at the graph. However this defining expression is already useful in order to derive differential equations satisfied by the generating function. Indeed, let us come back shortly to the case of the 2-vertex graph, for which we are left with a single integration over $\SU(2)$:
\be
\cW(w_i,z_i)=\int_{\SU(2)} dh\,
e^{\sum_i^N[z_i|h|w_i\ra}\,.
\ee
As we have already seen previously, tt is then direct to derive the second differential equations \eqref{wdw} characterizing the generating function:
\be
\left[ \partial_{z_i}|\partial_{z_j}\right\rangle \,\cW(w_i,z_i)
\,=\,
\int dh\,
-[w_j|h^{-1}h|w_i\ra
\,
e^{\sum_i^N[z_i|h|w_i\ra}
\,=\,
[w_i|w_j\rangle\,\cW(w_i,z_i)\,.
\ee
We can generalize this to arbitrary graph by choosing a loop of $\Gamma$, i.e. a closed sequence of edges and vertices  $v_1\overset{e_1}{\arr}v_2\dots v_n\overset{e_n}{\arr}v_1$, as shown on figure \ref{loop}.
The strategy already introduced in \cite{recursion}, and further developed in \cite{recursion3d,recursion4d,recursion_spinor,sfcosmo}, is to write a equation on the scalar product which is true on flat connection and to have it act on the spin network function. For instance, in our case, we consider the scalar product between the spinors at the origin vertex $v_1$ of the loop:
\be
\left.([z_n|g_n..g_1|w_1\ra-[z_n|w_1\ra)\right|_{g_i=\id}=0
\,\quad\longrightarrow\quad\,
\left.\left([z_n|\prod_i\widehat{g_i}|w_1\ra-[z_n|w_1\ra\right)\,
\varphi_{\Gamma,z^v_e}(g_e)\right|_{g_e=\id}=0
\ee
The scalar product $[z_n|w_1\ra$ acts by multiplication, but the action of the holonomy operators $\widehat{g_e}$ is non-trivial. They lead to shifts of the spins $j_e$ leaving on the corresponding edges in the standard spin network basis, and this action can be translated to differential operators in the spinor variables when acting on the presently used coherent spin network states.
This technique was used in \cite{recursion,recursion3d,recursion4d,recursion_spinor} to generate recursion relations on the 6j symbols and other 3nj symbols and to derive the Hamiltonian constraint of topological quantum gravity in 3d and 4d in the spinfoam framework. It was similarly used in \cite{sfcosmo} to derive the Wheeler-de-Witt Hamiltonian constraint for the (modified) Friedmann-Robertson-Walker homogeneous cosmology from the spinfoam amplitudes of the 2-vertex model for cosmology in loop quantum gravity.

Here too we can be completely explicit and write the full differential equations satisfied b the coherent spin network evaluation $\cW(z^v_e)$.
Similarly to the case of the 2-vertex graph, we differentiate with the operator $[ \partial_{z_i}|\partial_{z_j}\ra$ all around the loop except at the first vertex.
Starting the differentials at the vertices $v_2$ and $v_3$, we get:
\beq
[\partial_{z_2}|\partial_{w_3}\ra[\partial_{z_1}|\partial_{w_2}\ra\,\cW(z^v_e)
&=&
[\partial_{z_2}|\partial_{w_3}\ra\int [dh]^V\,
[z_2|h_3^{-1}h_1|w_1\ra\,
e^{[{z_1}|h_2^{-1}h_1|{w_1}\ra+[{z_2}|h_3^{-1}h_2|{w_2}\ra+\dots}\nn\\
&=&
\int [dh]^V\,
\left([z_3|h_4^{-1}h_1|w_1\ra
+[z_3|h_4^{-1}h_2|w_2\ra[z_2|h_3^{-1}h_1|w_1\ra
\right)\,
e^{[{z_1}|h_2^{-1}h_1|{w_1}\ra+[{z_2}|h_3^{-1}h_2|{w_2}\ra+\dots}\,.\nn
\eeq
Differentiating as such, we see that we slowly compose the holonomies as wanted but we also generate ``bad" terms. This is expected since we are acting only with the anti-holomorphic part of the holonomy operator. As explained \cite{spinor} and carried out explicitly in \cite{recursion_spinor} for the 6j symbol on the tetrahedral graph, we need to include all the holomorphic and anti-holomorphic contributions to the holonomy operator. This translates to:
\be
\bigg{[}
\big{(}[\partial_{z_1}|\partial_{w_2}\ra\big{)}\big{(}[\partial_{z_2}|\partial_{w_3}\ra\big{)}
-\big{(}[\partial_{z_1}|[w_2|\big{)}\big{(}|z_2\ra|\partial_{w_3}\ra\big{)}
\bigg{]}\,
\cW(z^v_e)
\,=\,
\int [dh]^V\,
[{z_2}|h_3^{-1}h_2|{w_2}\ra
\,
[z_3|h_4^{-1}h_1|w_1\ra
\,
e^{[{z_1}|h_2^{-1}h_1|{w_1}\ra+\dots}\,,\nn
\ee
after straightforward algebraic manipulations\footnote{We use the matrix identity on spinors $|w\ra[z|-|z\ra[w|=[z|w\ra\id$.}. Here the notation $|z\ra|\pp_w\ra$ means the contraction over spinor indices,
\be
|z\ra|\pp_w\ra\,\equiv\,
\tr  |z\ra \otimes|\pp_w\ra
=\sum_A z^A \pp_{w^A}
=[z|[\pp_w|\,,
\ee
We see in the previous calculation that we get the correct factor $[z_3|h_4^{-1}h_1|w_1\ra$ with the holonomy insertion along the loops times another factor $[{z_2}|h_3^{-1}h_2|{w_2}\ra+1$. $[{z_2}|h_3^{-1}h_2|{w_2}\ra$ is a term in the exponential and can be simply generated by differentiating the integral $\cW(z^v_e)$ with respect to the norm of $w_2$ (or equivalently $z_2$). In the spin representation, the factor $[{z_2}|h_3^{-1}h_2|{w_2}\ra+1$ is the dimension $(2j_2+1)$ of the representation carried by the link. We refer to \cite{recursion_spinor} for the precise relation between the partial differential equations satisfied in the coherent spinorial basis and the recursions satisfied in the spin representation.
Repeating this process all around the loop gives us finally a beautiful differential equation satisfied by the coherent spin network evaluation:
\be
[\partial_{z_1}|\,
\prod_{i=2}^{n-1}\big{(}
|\partial_{w_i}\ra\otimes[\partial_{z_i}|-
[w_i| \otimes|z_i\ra
\big{)}\,
|\partial_{w_n}\ra\,
\cW(z^v_e)
\,=\,
[z_n|w_1\ra\,
\prod_{i=2}^{n-1}\big{(}1+
|w_i\ra\otimes|\partial_{w_i}\ra
\big{)}\,
\cW(z^v_e)\,.
\ee
This equation (actually its dual, where multiplication and derivation are exchanged) has already been derived for a cycle of three links, $n=3$, in the appendix of \cite{recursion_spinor}.

An open question is whether this set of differential equations for all the loops of a given graph $\Gamma$ fully determine the coherent spin network evaluation $\cW(z^v_e)$ like in the case of the 2-vertex graph. For this purpose, it might be interesting to get an expression of the functions $\cW(z^v_e)$ directly in terms of the  variables $F$'s. This is a hard task that we postpone to future investigation.

\subsection{Computing the Generating Functions}


At the end of the day, we do not know the solutions to the above partial differential equations. However, the formal expression of $\cW(z_e^v)$ in terms of integrals over $\SU(2)$ enables to relate it to the Schwinger-type of coherent spin network evaluations which are easier to calculate. Those coherent spin networks are defined with different combinatorial weights than $\cW$ (taking for example $f_J=(J+1)!^2$ in the case of the 2-vertex graph) depending on the total area around each node of the graph. To account for these combinatorial weights, we need a different set of coherent intertwiners,
\be
|\{ z_i\}\ra^{alg}
\,=\,
\sum_J \sqrt{\f{(J+1)!}{J!}}\,|J,\{ z_i\}\ra
\,=\,
\sum_{j_i}\f{(J+1)!}{\sqrt{ \prod_i (2j_i)!}}\,|\{ j_i,z_i\}\ra\,.
\ee
We define the corresponding algebraic spin network function $\varphi_{\Gamma,z^v_e}^{alg}(g_e)$ and its evaluation $\cW(z^v_e)^{alg}\equiv\varphi_{\Gamma,z^v_e}^{alg}(\id)$. As shown in \cite{jeff}, we can use the result \ref{gaussian_result} of section \ref{generating_section} to switch the integrals over $\SU(2)$ into Gaussian integrals over spinor variables due to the $(J+1)!$ factors in the intertwiners. This leads to express this generating function {\it\`a la Schwinger} as a Gaussian integral:
\be
\varphi_{\Gamma,z^v_e}^{alg}(g_e)
\,=\,
\int_{\C^{2V}} \prod_v\f{[d^4\zeta_v]e^{-\la\zeta_v|\zeta_v\ra}}{\pi^2}\,
e^{\sum_e[z_e|H_{t(e)}^{-1}g_eH_{s(e)}|w_e\ra}\,,
\ee
with $H_v\equiv \big{(}|\Omega\ra[\zeta_v|-|\Omega]\la\zeta_v|\big{)}$. Let us insist on the fact that these $H_v$'s are not $\SU(2)$ group element since the spinors $\zeta_v$ are not normalized.
This formula simplifies for the evaluation at the identity:
\be
\cW(z^v_e)^{alg}
\,=\,
\int_{\C^{2V}} \prod_v\f{[d^4\zeta_v]e^{-\la\zeta_v|\zeta_v\ra}}{\pi^2}\,
e^{\sum_e[z_e|\zeta_{t(e)}\ra \la\zeta_{s(e)}|w_e\ra+[z_e|\zeta_{t(e)}][\zeta_{s(e)}|w_e\ra}\,.
\ee
Since this is a complex Gaussian integral, it is possible to compute it exactly. This was precisely done in \cite{jeff}, where the authors gave a pretty expression of $\cW(z^v_e)^{alg}$ as a rational function in the spinors $z^v_e$ (or more explicitly in the product of the $F$'s along loops of the graph). We will not repeat these results  here and we refer the interested reader to \cite{jeff} for all the details.
The idea which we wish to expand upon here is that it should be possible to compute the coherent spin network evaluation $\cW(z^v_e)$ from the Schwinger generating function $\cW(z^v_e)^{alg}$, as we have shown for the 2-vertex graph in section \ref{compute}. Indeed the integral over $\SU(2)$ can be directly translated into an integral over {\it normalized} spinors:
\be\nn
\cW(z^v_e)
\,=\,
\int_{\C^{2V}} \prod_v\f{[d^4\zeta_v]}{\pi^2}\,
\delta(\la\zeta_v|\zeta_v\ra-1)\,
e^{\sum_e[z_e|\zeta_{t(e)}\ra \la\zeta_{s(e)}|w_e\ra+[z_e|\zeta_{t(e)}][\zeta_{s(e)}|w_e\ra}\,.
\ee
We can Fourier transform the $\delta$-distribution and write\footnote{
In order to keep the integrals well-defined, it is certainly better to Fourier transform the spherical constraints using the exact identity $2\pi\,\delta(x)=\int dT \exp((iT-\eps)x)$ for all choices of $\eps$. This leads to the slightly modified Gaussian integrals
\be
\cW(z^v_e)
\,=\,
\int_{\R^V} \prod_v e^{\eps-iT_v}\f{dT_v}{2\pi}
\int_{\C^{2V}} \prod_v\f{[d^4\zeta_v]e^{-(\eps-iT_v)\la\zeta_v|\zeta_v\ra}}{\pi^2}\,
e^{\sum_e[z_e|\zeta_{t(e)}\ra \la\zeta_{s(e)}|w_e\ra+[z_e|\zeta_{t(e)}][\zeta_{s(e)}|w_e\ra}\,,
\ee
with a clearly better behaved Hessian matrix on the $\zeta$'s for an arbitrary positive value for $\eps>0$.}
\be
\cW(z^v_e)
\,=\,
\int_{\R^V} \prod_v e^{-iT_v}\f{dT_v}{2\pi}
\int_{\C^{2V}} \prod_v\f{[d^4\zeta_v]e^{iT_v\la\zeta_v|\zeta_v\ra}}{\pi^2}\,
e^{\sum_e[z_e|\zeta_{t(e)}\ra \la\zeta_{s(e)}|w_e\ra+[z_e|\zeta_{t(e)}][\zeta_{s(e)}|w_e\ra}\,.
\ee
Putting aside the integrals over the ``proper time variables" $T_v$, this is very similar to the definition of Schwinger generating function $\cW(z^v_e)^{alg}$ up to the $iT_v$ factor in the Gaussian measure over the $\zeta_v$ variables. It seems to us that it is thus possible to integrate entirely over the spinors $\zeta_v$, either by a suitable rescaling of $\cW(z^v_e)^{alg}$ or by repeating the same Gaussian integral as in \cite{jeff} taking into account those $iT_v$ factors. Then $\cW(z^v_e)$ would simply be some Fourier transform of some rescaling of the function $\cW(z^v_e)^{alg}$.

Let us finish this discussion by stating that it is possible to derive a differential equation for $\cW(z^v_e)^{alg}$ following the analysis carried out above for $\cW(z^v_e)$. We will now have extra factors due to the fact that the spinors $\zeta_v$ are not normalized, which will complicate the expression of the differential operators but which will not modify deeply its structure.

We leave a detailed study of the properties of those generating functions $\cW(z^v_e)$ and $\cW(z^v_e)^{alg}$ on arbitrary graphs for future investigation.

\section*{Outlook}

The main results of the paper are
\begin{itemize}
 \item a formula for the scalar product of coherent intertwiners at fixed spins and cross-ratios,
 \item the \emph{exact} calculation of generating functions for spin network evaluations on the 2-vertex graph, with \emph{different} possible choices of combinatorial weights, including Schwinger's choice \cite{schwinger:52} and another one we have called the \emph{geometric} one,
 \item the geometrical interpretation of their saddle point evaluations,
 \item the Wheeler-DeWitt equation for $\SU(2)$-flat dynamics on arbitrary graphs (it has the simplest form for the geometric generating function).
\end{itemize}
Therefore this work strongly support the idea that loop quantum gravity and spin foams would take advantage of being re-formulated in terms of spinors and Schwinger's bosonic operators. In particular in spin foam models, amplitudes are given by sums over spins of products of spin network evaluations. Hence the generating function methods we have introduced here are likely to apply. We expect that such a formulation will make the tools of (complex) analysis applicable which would be a great improvement compared to sums over spins of complicated Wigner symbols.

While the (mathematical) literature has focused so far on the Schwinger's choice of combinatorial weights, we have shown other choices are possible and are further maybe more interesting for physical reasons, in particular the geometric generating function. But it requires some work to see how such generating function can be evaluated for arbitrary graphs, beyond our simple 2-vertex example. In the case of arbitrary graphs, we have shown that the geometric function can be recast as a Fourier transform of a suitably rescaled Schwinger's generating function and derived a set of partial differential equations. We hope those features will help us evaluate the geometric generating function in the future. We believe that this is the function which will always have a natural geometric interpretation in its saddle point evaluation and admit the most natural form of Wheeler-DeWitt equation, at least for simple dynamics.

Another natural extension we have pointed out is the insertion of non-trivial Wilson lines to introduce curvature in a simple way. However further work is needed to write partial differential equations satisfied by this function.

\section*{Acknowledgments}

The authors are thankful to Laurent Freidel and Jeff Hnybida for sharing their results and discussions on resummations of spin network evaluations.

Research at Perimeter Institute is supported by the Government of Canada through Industry Canada and by the Province of Ontario through the Ministry of Research and Innovation.

\appendix



\section{Free and Constrained Gaussian Integrals}
\label{gaussian}

We compare the evaluation of the usual two-dimensional Gaussian integral $I\equiv\int d^2x\, \exp(-{}^txMx)=2\pi/\sqrt{\det M}$ and the constrained Gaussian integral $\tl{I}\equiv\int d^2x\,\delta(x^2-1)\, \exp(-{}^txMx)$ where the vector is constrained to be of unit norm.
For the sake of simplicity, we consider the special case where the matrix $M$ is real and symmetric and defined in terms of two parameters $a$ and $b$:
\be\nn
M=\mat{cc}{a & b \\ b & a},\qquad\textrm{with}\quad |b|<a\,.
\ee
The Gaussian integral can be easily evaluated with different methods. Here we go to radial coordinates:
\be
I=\int_{\R^2} dxdy\,e^{-\f12(ax^2+ay^2+2bxy)}
=\int_0^{2\pi}d\theta\int_0^{+\infty}dr\,
re^{-\f12r^2(a+b\sin2\theta)}
=\int_0^{2\pi}d\theta\,\f1{a+b\sin2\theta}
=\f{2\pi}{\sqrt{a^2-b^2}}\,.
\ee
We compare to the evaluation of the constrained Gaussian integral:
\be
\tl{I}=
\int_0^{2\pi}d\theta\int_0^{+\infty}dr\,\delta(r^2-1)\,re^{-\f12r^2(a+b\sin2\theta)}
=\f{e^{-\f a2}}2\int_0^{2\pi}d\theta\,e^{-\f b2\sin2\theta}
=\pi\,e^{-\f a2}\,I_0\left(\f b2\right)\,,
\ee
given by the modified Bessel function $I_0$. This evaluation does not actually involve any Gaussian integral. We can nevertheless compute it by Fourier transforming the constraint and performing the Gaussian integral:
\be
\tl{I}=
\f1{2\pi}\int_{\R}dT\int dxdy\,
e^{-iT}\,e^{-\f12(a-2iT)(x^2+y^2)}e^{-bxy}
=
\int_{\R}dT\,\f{e^{-iT}}{\sqrt{(a-2iT)^2-b^2}}
=
\pi\,e^{-\f a2}\,I_0\left(\f b2\right)\,.
\ee

\section{The Generating Functions and Wigner 3nj-Symbols} \label{sec:GF-wigner3nj}

Generating functions for 3nj-symbols considered in the past are based on 3-valent coherent intertwiners. As we use in this paper $N$-valent nodes, we would like to know whether our generating functions generate Wigner 3nj-symbols too.

\subsection{Spin Network Basis and 3nj-Symbols}

\begin{figure}[h]
\begin{center}
\includegraphics[height=30mm]{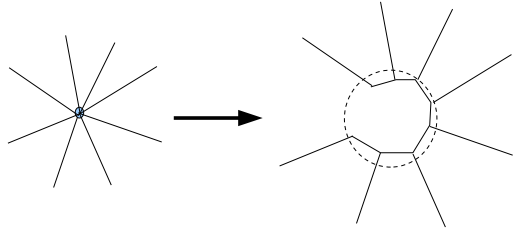}
\caption{A 3-valent tree unfolding of the N-valent vertex for here $N=8$. There always exist many unfoldings of the vertex for $N\ge 4$.\label{intertwiner_basis}}
\end{center}
\end{figure}

Wigner symbols arise as change of basis between different specific bases of intertwiners. Standard representation theory of $\SU(2)$ states that, due to the uniqueness of the Clebsh-Gordan coefficients, a basis of the intertwiner space for fixed external spins $j_1,..,j_N$ is constructed by choosing a 3-valent tree unfolding of the $N$-valent vertex and label the internal edges of the tree with spins. More precisely, we choose a graph with no loop (tree) whose vertices are all 3-valent and with $N$ external legs, as illustrated in fig.\ref{intertwiner_basis}.
Such a 3-valent tree contains $(N-3)$ edges and thus a basis state for the intertwiner space will be labeled by $(N-3)$ spins. Therefore, on the 2-vertex graph, one gets an orthonormal basis of spin network states labeled by $N$ spins $j_i\in\N/2$ living on the edges, $(N-3)$ spins $j^\alpha_k$ to define the intertwiner $i_\alpha$ and $(N-3)$ spins $j^\beta_k$ to define the intertwiner $i_\beta$, which means $3(N-2)$ spins in total.
For a given set of $N$ external legs, there are many ways to choose such 3-valent trees, leading to different bases of the intertwiner space, and there is no need to choose the same 3-valent unfolding for both vertices $\alpha$ and $\beta$.

By definition, scalar products between intertwiners with different choices of trees give rise to Wigner 3nj-symbols, which are the objects of study of $\SU(2)$ re-coupling theory. The 2-vertex graph is interesting in this respect because the evaluation of a spin network on the identity, $\vphi_{j_i,i_\alpha,i_\beta}(g_i=\id)$, is exactly the scalar product between the intertwiners $i_\alpha$ and $i_\beta$ in the Hilbert space $\textrm{Inv}_{\SU(2)} \bigotimes_i \cV^{j_i}$,
\be
\vphi_{j_i,i_\alpha,i_\beta}(g_1,..,g_N)
\,=\,
\la i_\alpha | \otimes D^{j_i}(g_i)|i_\beta\ra,
\qquad
\vphi_{j_i,i_\alpha,i_\beta}(\id)
\,=\,
\la i_\alpha | i_\beta\ra\,.
\ee
When unfolding the two vertices into 3-valent trees, $\cT^\alpha$ and $\cT^\beta$, and labeling the intertwiners with spins, $j^\alpha_k$ and $j^\beta_k$, the scalar products can produce all 3nj-symbols with $n\leq(N-2)$. Indeed, taking the same tree on both sides $\cT^\alpha=\cT^\beta$ is the trivial case where the scalar product simply reflects the orthonormality of the basis and identifies the two intertwiners, $\prod_k^{N-3}\delta_{j^\alpha_kj^\beta_k}$. On the other hand, considering different trees for the two vertices provide us with all non-trivial 3nj-symbols.

\begin{figure}[h]
\begin{center}
\includegraphics[height=30mm]{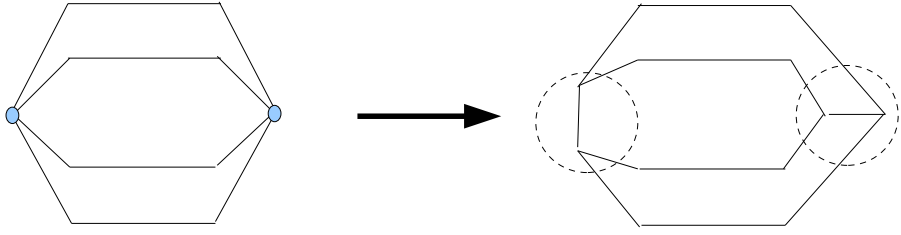}
\caption{The 6j-symbol as the evaluation of the $N=4$ spin network states on the identity: we unfold the two vertices $\alpha$ and $\beta$ into two different 3-valent trees, this gives the tetrahedral graph with four vertices.\label{6j_fig}}
\end{center}
\end{figure}

For instance, for $N=4$ edges, 3-valent trees contain a single internal link and we recover the standard definition of the 6j-symbol ($n=N-4=2$) when $\cT^\alpha$ and $\cT^\beta$ are different. It is the scalar product between basis states for the two different bases for re-coupling 3 spins together, for example re-coupling $j_1\otimes j_2\otimes j_3$ into $j_4$ as $((j_1\otimes j_2)\otimes j_3)\arr j_4$ or as $(j_1\otimes (j_2\otimes j_3))\arr j_4$ as shown in the figure \ref{6j_fig}.

\subsection{Relations between the Generating Functions and Wigner 3nj-symbols}

Further decomposing the LS intertwiners $|\{j_i,w_i\}\ra$ and $|\{j_i,z_i\}\ra$ by unfolding the vertices $\alpha$ and $\beta$ with appropriate 3-valent trees $\mathcal T^\alpha, \mathcal T^\beta$ and internal spins $j^\alpha_k, j^\beta_l$, we get
\be \label{pre-generating}
\cW^{f_J}(w_i,z_i)
=
\sum_{\{j_i,j^\alpha_k,j^\beta_l\}} \f{f_{\sum_ij_i}}{\prod_i (2j_i)!}\,\la \{j_i,\varsigma z_i\}|\{j_i,j^\alpha_k\}\ra\,\la \{j_i,j^\beta_l\}|\{j_i,w_i\}\ra\, \la \{j_i,j^\alpha_k\}|\{j_i,j^\beta_l\}\ra\,.
\ee
We see that $\cW(w_i,z_i)$ is a sum over spins of $\SU(2)$ 3nj-symbols, denoted here $\la \{j_i,j^\alpha_k\}|\{j_i,j^\beta_l\}\ra$, with appropriate polynomials in $z_i$ and $w_i$, respectively $\la \{j_i,\varsigma z_i\}|\{j_i,j^\alpha_k\}\ra$ and $\la \{j_i,j^\beta_l\}|\{j_i,w_i\}\ra$. Moreover, these polynomials are homogeneous holomorphic polynomials of degree $j_i$. That suggests to interpret $\cW(w_i,z_i)$ as a generating function for 3nj-symbols. However, this is more subtle. Indeed, there is no variable {\it a priori} to control the internal spins $j^\alpha_k,j^\beta_l$. Further, $\cW$ should be expressed in terms of $\SU(2)$ invariant variables to understand really what objects it generates.

Introducing the factorization of coherent intertwiners of the type \eqref{factor coherent state} into \eqref{pre-generating} leads to
\be \label{generating function hol brackets}
\cW^{f_J}(w_i,z_i) = \sum_{\{j_i,j^\alpha_k,j^\beta_l\}} \f{f_{\sum_ij_i}}{\prod_i (2j_i)!}\,[\prod_{ij} F_{ij}(z)^{\Delta^\alpha_{ij}}] [\prod_{ij} F_{ij}(w)^{\Delta^\beta_{ij}}]\ \la \{j_i,Z_i\}|\{j_i,j^\alpha_k\}\ra\, \la \{j_i,j^\alpha_k\}|\{j_i,j^\beta_l\}\ra\, \la \{j_i,j^\beta_l\}|\{j_i,W_i\}\ra\,.
\ee
Here the products run over some independent sets of $\SU(2)$ invariant variables, like \eqref{su2 invariant var} and its permutations between between links. The exponents $\Delta^\alpha_{ij}, \Delta^\beta_{ij}$ are read on \eqref{factor coherent state} and permutations.

One can go further with the knowledge of the brackets $\la \{j_i,j^\beta_l\}|\{j_i,W_i\}\ra$. However, they have only been studied in the simplest non-trivial case $N=4$, in \cite{holquantumtet}. Assuming that the generic case is also polynomial in the cross-ratios, we can write
\be \label{polynome expansion}
\la \{j_i,j^\beta_l\}|\{j_i,W_i\}\ra = \sum_{\{q^\beta_l\}} P_{q^\beta_l}(j_i,j^\beta_l)\, \prod_{l=4}^N W_l^{q^\beta_l}\,.
\ee
Then we obtain
\be \label{generating function 3nj}
\cW^{f_J}(w_i,z_i) = \sum_{\{j_i,p^\alpha_k,q^\beta_l\}} \f{f_{\sum_ij_i}}{\prod_i (2j_i)!}\,[\prod_{ij} F_{ij}(z)^{\Delta^\alpha_{ij}}] [\prod_{ij} F_{ij}(w)^{\Delta^\beta_{ij}}] [\prod_{k} Z_k^{p^\alpha_k}][\prod_{l} W_l^{q^\beta_l}]\ \cW(j_i,p^\alpha_k,q^\beta_l)\,.
\ee
Hence, $\cW^{f_J}$ is a generating function for the following objects
\be
\cW(j_i,p^\alpha_k,q^\beta_l) = \sum_{\{j^\alpha_k,j^\beta_l\}} P_{p^\alpha_k}(j_i,j^\alpha_k)\ P_{q^\beta_l}(j_i,j^\beta_l)\ \la \{j_i,j^\alpha_k\}|\{j_i,j^\beta_l\}\ra\,.
\ee
Such an object is a sum of 3nj-symbols at fixed spins $j_i$ weighted by the coefficients of the expansion \eqref{polynome expansion}. Just below we give some details on the (somewhat trivial) case $N=3$ and the (more interesting) case $N=4$.

\subsubsection{The 3-valent case}

The set of intertwiners, i.e. vectors in ${\rm Inv} (\cV^{j_1}\otimes \cV^{j_2}\otimes \cV^{j_3})$, is then one-dimensional, spanned by the normalized state $|j_1 j_2 j_3\ra$ whose components in the usual magnetic number basis are the Wigner 3jm-symbols
\be
\la \{j_1,j_2,j_3 ; m_1, m_2, m_3\}| j_1 j_2 j_3\ra = \begin{pmatrix} j_1 &j_2 &j_3\\m_1 &m_2 &m_3\end{pmatrix}\,.
\ee
Therefore, the coherent LS intertwiner is proportional to $| j_1 j_2 j_3\ra$,
\be
|\{j_1,j_2,j_3;z_1,z_2,z_3\}\ra = P_{j_1 j_2 j_3}(z_1,z_2,z_3)\ | j_1 j_2 j_3\ra\,,
\ee
where $P_{j_1 j_2 j_3}$ is the generating function for the 3jm-symbols with fixed spins,
\be
P_{j_1 j_2 j_3}(z_1,z_2,z_3) = \prod_{e=1}^3 \sqrt{(2j_e)!}\f{(z_e^1)^{j_e-n_e}\ (z_e^0)^{j_e+n_e}}{\sqrt{(j_e-n_e)!\,(j_e+n_e)!}}\,\begin{pmatrix} j_1 &j_2 &j_3\\ n_1 &n_2 &n_3\end{pmatrix}\,,
\ee
and admit a closed formula as an invariant holomorphic polynomial \cite{varshalovich},
\be
\begin{aligned}
P_{j_1 j_2 j_3}(z_1,z_2,z_3) &=\frac{\sqrt{(2j_1)!\,(2j_2)!\,(2j_3)!}}{\Delta(j_1 j_2 j_3)\,(J+1)!}\ F_{12}(z)^{J-2j_3}\ F_{23}(z)^{J-2j_1}\ F_{31}(z)^{J-2j_2}\,,\\
\Delta(j_1 j_2 j_3) &= \sqrt{\frac{(J-2j_1)!\,(J-2j_2)!\,(J-2j_3)!}{(J+1)!}}\,.
\end{aligned}
\ee
There is obviously no cross-ratio, because there is no Pl\"{u}cker relation between $F_{12}, F_{23}, F_{31}$ which are independent variables. Hence, the coherent spin network evaluation is simply
\be
\begin{aligned}
\cW^{f_J}(z_i,w_i) &= \sum_{j_1,j_2,j_3} \frac{f_{\sum_i j_i}}{\prod_{i=1}^3 (2j_i)!}\ P_{j_1,j_2,j_3}(z_1,z_2,z_3)\ P_{j_1,j_2,j_3}(w_1,w_2,w_3)\,,\\
&= \sum_{j_1,j_2,j_3} \frac{f_{\sum_ij_i}}{\Delta(j_1 j_2 j_3)^2\,(J+1)!^2}\ \bigl[F_{12}(z)F_{12}(w)\bigr]^{J-2j_3}\, \bigl[F_{23}(z)F_{23}(w)\bigr]^{J-2j_1}\, \bigl[F_{13}(z)F_{13}(w)\bigr]^{J-2j_2}\,.
\end{aligned}
\ee
This is the expected form of the generating function \cite{varshalovich}, which however only generates factorials depending on $j_1,j_2,j_3$.

\subsubsection{The 4-valent case}

In the standard spin network basis, specifying the intertwiner requires a tree expansion of the 4-valent node. There are three possible channels. For instance, the channel $(12)$ corresponds to a tree which has a node where the links 1 and 2 meet, and another node where the links 3 and 4 meet. The two nodes are connected by a line and one has to choose a spin $j$ on this line to completely determine the intertwiner $|\{j_i\},j\ra^{12}$. The other channels are obtained by exchanging the link 1 with the link 3 or 4.

In the coherent basis, there is one cross-ratio per intertwiner, which we denote $Z_4$ and $W_4$, and define like in \eqref{def cross-ratio}. With our choice of cross-ratios, the channel $(12)$ is somehow distinguished by the fact that the bracket ${}^{12}\la \{j_i\},j^\beta|\{j_i\},W_4\ra$ satisfies a hypergeometric equation \cite{holquantumtet}. As a result, it takes the form
\be \label{jacpoly}
{}^{12}\la \{j_i\},j^\beta|\{j_i\}, W_4\ra = \sum_{p=0}^{j^\beta-j_3+j_4} P_p(j_i,l)\,W^p_4\,,
\ee
which is proportional to some shifted Jacobi polynomial\footnote{In details,
\begin{multline}
{}^{12}\la \{j_i\},j^\beta|\{j_i\}, W_4\ra = (-1)^{J-2j^\beta}\sqrt{\frac{\prod_{i=1}^4 (2j_i)!\, (j^\beta+j_3-j_4)! (j^\beta+j_4-j_3)!}{(j_1+j_2+j^\beta+1)! (j_3+j_4+j^\beta+1)! (j_1+j_2-j^\beta)! (j_3+j_4-j^\beta)! (j^\beta+j_1-j_2)! (j^\beta+j_2-j_1)!}}\\
\sqrt{2j^\beta+1}\ P^{(j_3-j_4-j_1+j_2,j_3-j_4+j_1-j_2)}_{j^\beta -j_3+j_4}(1+2W_4)\,,
\end{multline}
where $P^{(a,b)}_n(z)$ is the Jacobi polynomial.}.

We first choose the same channel $(12)$ for the two nodes of the graph. With our choice of normalization, ${}^{12}\la \{j_i\},j^\alpha|\{j_i\},j^\beta\ra^{12} =\delta_{j^\alpha,j^\beta}$. Coming back to \eqref{generating function 3nj}, we thus get
\begin{multline} \label{N=4 generating}
\cW(w_i,z_i)
=
\sum_{\{j_i\},p,q}[F_{12}(w)F_{12}(z)]^{J-2j_3} [F_{13}(w)F_{13}(z)]^{2j_1+2j_3-J} [F_{23}(w)F_{23}(z)]^{2j_2+2j_3-J} [F_{43}(w)F_{43}(z)]^{2j_4}\\
\times\, W_4^p\,Z_4^q\ \f1{\prod_i (2j_i)!}\biggl[\sum_{j} P_p(j_i,j) P_q(j_i,j)\biggr]\,.
\end{multline}
For any fixed set of spins $\{j_i\}$, the remaining sums over $p,q,j$ are finite (though we do not write their range to avoid cumbersome expressions). This is the form of the generating function, as a function of the $F_{12}, F_{13}, F_{23}, F_{43}, W_4, Z_4$.

One can actually go further by comparing the above formula with the expansion \eqref{GF} of $\cW$, and identifying their coefficients. To avoid too large formulas, we will illustrate this in the case $N=4$ again, which gives
\begin{multline}
\f{(1+\sum_i j_i)!}{\prod_i (2j_i)!}\ \sum_{l} P_m(j_i,l) P_n(j_i,l)
=
\sum_{l\geq m,n}^{2j_4}\sum_{p_{24}=0}^{l-m,l-n}\frac{p_{24}!}{(j_1+j_2+j_{34}-l)! (j_{34}+j_{12}+p_{24})! (j_{34}-j_{12}+l-p_{24})!}\\
\frac1{(2j_4-l)! (p_{24}-l+n)! (p_{24}-l+m)! (l-m)! (l-n)!} \,.
\end{multline}
We have used the notation $j_{ij}\equiv j_i-j_j$. This relation holds for any fixed $m,n$. Interestingly, the sums over $l$ and $p_{24}$ decouple when $j_1=j_2, j_3=j_4$, and can be performed explicitly. In addition, coefficients $P_m(j_i,l)$ of Jacobi polynomials on the left hand side reduce to coefficients of the Legendre polynomials. That gives the following equality
\be
\sum_l \frac{(2l+1)\ (2j_1+2j_4+1)!}{(2j_1+l+1)! (2j_1-l)! (2j_4+l+1)! (2j_4-l)!} \biggl[\frac{(l+m)! (l+n)!}{(l-m)! (l-n)!}\biggr]
=
\frac{(2j_1+2j_4-m-n)! (m+n)!}{(2j_1-m)! (2j_1-n)! (2j_4-m)! (2j_4-n)!}\,.
\ee

Notice that such expansions can not yield directly the coefficients $P_{p^\alpha_k}(j_i,j^\alpha_k)$ of \eqref{polynome expansion} (and the 3nj-symbols with all spins fixed neither), because there is no way here to get the internal spins $j^\alpha_k$ fixed. Instead, our expansions provide a formula for $\sum_{j^\alpha_k,j^\beta_l} P_{p^\alpha_k}(j_i,j^\alpha_k) P_{q^\beta_l}(j_i,j^\beta_l)\prod_k \delta_{j^\alpha_k,j^\beta_k}$, at fixed spins $j_i$ and fixed $p^\alpha_k,q^\beta_l$ which are the exponents of the cross-ratios, like in \eqref{N=4 generating}. Those formulas can be seen as sum rules for products of coefficients of Jacobi polynomials.

One can also consider a different channel for, say, the node $\alpha$, like the channel $(23)$. However, the bracket $\la \{j_i\},Z_4|\{j_i\},j^\alpha\ra^{23}$ is not a Jacobi polynomial anymore. It is thus more convenient to choose a more adapted cross-ratio which would make the bracket a Jacobi polynomial. As shown in \cite{holquantumtet}, that is obtained by exchanging the lines 1 and 3 everywhere, which is indeed quite natural since exactly this characterizes the change of channels from $(12)$ to $(23)$. Then we have \cite{holquantumtet}
\be
| \{j_i,z_i\}\ra = F_{23}(z)^{J-2j_1} F_{13}(z)^{2j_1+2j_3-J} F_{12}(z)^{2j_1+2j_2-J} F_{14}(z)^{2j_4}\ |\{j_i\},1/Z_4\ra\,,
\ee
for which the bracket $\la \{j_i\},1/Z_4|\{j_i\},j^\alpha\ra^{23}$ is basically a Jacobi polynomial in $1/Z_4$ and hence has an expansion like \eqref{jacpoly}. When choosing the channels $(12)$ on the node $\beta$ and $(23)$ on $\alpha$, the bracket ${}^{23}\la \{j_i\},j^\alpha| \{j_i\},j^\beta\ra^{12}$ is by definition the Wigner 6j-symbol. Hence, we now obtain a different form of the coherent spin network evaluation, as a function of $F_{12}(z), F_{12}(w), F_{13}(z), F_{13}(w), F_{14}(z), F_{34}(w), F_{23}(z), F_{23}(w)$ and $W_4$ and $1/Z_4$,
\begin{multline}
\cW =
\sum_{\{j_i\},p,q}F_{12}(w)^{J-2j_3} F_{12}(z)^{2j_1+2j_2-J} [F_{13}(w)F_{13}(z)]^{2j_1+2j_3-J} F_{23}(w)^{2j_2+2j_3-J} F_{23}(z)^{J-2j_1} [F_{43}(w)F_{41}(z)]^{2j_4}\\
\times\, W_4^p\,(1/Z_4)^q\ \f1{\prod_i (2j_i)!}\biggl[\sum_{j^\alpha,j^\beta} P_p(j_i,j^\beta) P_q(j_i,j^\alpha) \begin{Bmatrix} j_1& j_2 &j^\beta\\j_3 &j_4 &j^\alpha\end{Bmatrix} \biggr]\,.
\end{multline}



\begin{thebibliography}{99}

\bibitem{twisted1}
L. Freidel and S. Speziale,
{\it Twisted geometries: A geometric parametrisation of SU(2) phase space},
Phys.Rev.D82 (2010) 084040 [arXiv:1001.2748]

\bibitem{polyhedron}
E. Bianchi, P. Dona and S. Speziale,
{\it Polyhedra in loop quantum gravity},
Phys.Rev.D83 (2011) 044035 [arXiv:1009.3402]

\bibitem{varshalovich}
  D.~A.~Varshalovich, A.~N.~Moskalev and V.~K.~Khersonsky,
  {\it Quantum theory of angular momentum: irreducible tensors, spherical harmonics, vector coupling coefficients, 3NJ symbols,}
{\it  Singapore, Singapore: World Scientific (1988) 514p}

\bibitem{spinnets-marzuoli}
  V.~Aquilanti, A.~C.~P.~Bitencourt, C.~d.~S.~Ferreira, A.~Marzuoli and M.~Ragni,
  {\it Quantum and semiclassical spin networks: From atomic and molecular physics
  to quantum computing and gravity,}
  Phys.\ Scripta {\bf 78}, 058103 (2008)
  [arXiv:0901.1074 [quant-ph]].

\bibitem{3nj-marzuoli}
  R.~W.~Anderson, V.~Aquilanti and A.~Marzuoli,
  {\it 3nj Morphogenesis and Semiclassical Disentangling,}
  J.\ Phys.\ Chem.\ {\bf A 113} (2009) 15106.
  arXiv:1001.4386 [quant-ph].

\bibitem{littlejohn}
V. Aquilanti, H.M. Haggard, A. Hedeman, N. Jeevanjee, R.G. Littlejohn and L. Yu,
{\it Semiclassical Mechanics of the Wigner 6j-Symbol},
J. Phys. A {\bf 45} (2012) 065209.
arXiv:1009.2811.

\bibitem{Yu}
  R.~G.~Littlejohn and L.~Yu,
  {\it Semiclassical Analysis of the Wigner $9J$-Symbol with Small and Large
  Angular Momenta,}
  Phys.\ Rev.\  A {\bf 83}, 052114 (2011)
  [arXiv:1104.1499 [math-ph]].\\
  L.~Yu,
  {\it Semiclassical Analysis of the Wigner $12J$-Symbol with One Small Angular Momentum: Part I,}
  Phys. Rev. A {\bf 84} (2011) 022101
  arXiv:1104.3275 [math-ph].\\
  L.~Yu,
  {\it Asymptotic Limits of the Wigner $15J$-Symbol with Small Quantum Numbers,}
  arXiv:1104.3641 [math-ph].
  L.~Yu,
  {\it Asymptotic Limits of the Wigner $12J$-Symbol in Terms of the Ponzano-Regge Phases,}
  arXiv:1108.1881 [math-ph].


\bibitem{3njsmall}
  V.~Bonzom, P.~Fleury,
  {\it Asymptotics of Wigner 3nj-symbols with Small and Large Angular Momenta: An Elementary Method,}
  J. Phys. A {\bf 45} (2012) 075202.
  [arXiv:1108.1569 [quant-ph]].

\bibitem{dowdall-handlebodies}
  R.~J.~Dowdall, H.~Gomes and F.~Hellmann,
  {\it Asymptotic analysis of the Ponzano-Regge model for handlebodies,}
  J.\ Phys.\ A  {\bf 43}, 115203 (2010)
  [arXiv:0909.2027 [gr-qc]].

\bibitem{6jnlo}
  V.~Bonzom, E.~R.~Livine, M.~Smerlak and S.~Speziale,
  {\it Towards the graviton from spinfoams: The Complete perturbative expansion of the 3d toy model,}
  Nucl.\ Phys.\  B {\bf 804}, 507 (2008)
  [arXiv:0802.3983 [gr-qc]].

\bibitem{pushing6j}
  M.~Dupuis and E.~R.~Livine,
  {\it Pushing Further the Asymptotics of the 6j-symbol,}
  Phys.\ Rev.\  D {\bf 80}, 024035 (2009)
  [arXiv:0905.4188 [gr-qc]].

\bibitem{6jmaite}
  M.~Dupuis and E.~R.~Livine,
  {\it The 6j-symbol: Recursion, Correlations and Asymptotics,}
  Class.\ Quant.\ Grav.\  {\bf 27}, 135003 (2010)
  [arXiv:0910.2425 [gr-qc]].

\bibitem{barrett-asym-summary}
  J.~W.~Barrett, R.~J.~Dowdall, W.~J.~Fairbairn, H.~Gomes, F.~Hellmann and R.~Pereira,
  {\it Asymptotics of 4d spin foam models,} (2010).
  arXiv:1003.1886 [gr-qc].

\bibitem{recursion6j_bis}
V. Bonzom and E.R. Livine,
{\it A New Recursion Relation for the 6j-Symbol},
Annales Henri Poincare (2011) [arXiv:1103.3415]

\bibitem{recursion}
V. Bonzom, E.R. Livine and S. Speziale,
{\it Recurrence relations for spin foam vertices},
Class. Quant. Grav. {\bf 27} (2010) 125002 [arXiv:0911.2204]

\bibitem{schwinger:52}
  J.~Schwinger,
  {\it On angular momentum,} Report US AEC NYO-3071 (1952),
  in Quantum Theory of Angular Momentum, eds. LC Biedenharn and H. van Dam (New York: Academic, 1965).

\bibitem{bargmann:62}
  V.~Bargmann,
  {\it On the Representations of the Rotation Group,}
  Rev.\ Mod.\ Phys.\  {\bf 34}, 829 (1962).

\bibitem{wu-9j}
  A.C.T.~Wu,
  {\it Structure of the Wigner 9j Coefficients in the Bargmann Approach,}
  J.\ Math.\ Phys.\ {\bf 13}, 84--90 (1972).

\bibitem{huang-wu-2j-15j}
  C.~-S.~Huang and A.~C T.~Wu,
  {\it Structure of the 12j and 15j coefficients in the bargmann approach,}
  J.\ Math.\ Phys.\  {\bf 15}, 1490 (1974).

\bibitem{labarthe}
  J.~J.~Labarthe,
  {\it Generating Functions for the Coupling Recoupling Coefficients of SU(2),}
  J.\ Phys.\ A  {\bf 8}, 1543 (1975).

\bibitem{schnetz}
  O.~Schnetz,
  {\it Generating Functions for Multi-j-Symbols,}
  [arXiv:math-ph/9805027].

\bibitem{garoufalidis}
  S.~Garoufalidis, R.~van der Veen and w.~a.~a.~Zagier,
  {\it Asymptotics of classical spin networks,}
  arXiv:0902.3113 [math.GT].

\bibitem{costantino-generating}
  F.~Costantino and J.~March\'e
  {\it Generating series and asymptotics of classical spin networks,}
  arXiv:1103.5644 [math.GT].

\bibitem{recursion_spinor}
V. Bonzom and E.R. Livine,
{\it A new Hamiltonian for the Topological BF phase with spinor networks},
arXiv:1110.3272

\bibitem{jeff}
  L.~Freidel and J.~Hnybida,
  {\it On the exact evaluation of spin networks,}
  arXiv:1201.3613 [gr-qc].

\bibitem{twisted2}
L. Freidel and S. Speziale,
{\it From twistors to twisted geometries},
Phys.Rev.D {\bf 82} (2010) 084041 [arXiv:1006.0199]

\bibitem{un0}
F. Girelli and E.R. Livine,
{\it Reconstructing Quantum Geometry from Quantum Information: Spin Networks as Harmonic Oscillators},
Class. Quant. Grav. {\bf 22} (2005) 3295-3314 [arXiv:gr-qc/0501075]

\bibitem{un1}
L. Freidel and E.R. Livine,
{\it The Fine Structure of SU(2) Intertwiners from U(N) Representations},
J. Math. Phys. {\bf 51} (2010) 082502 [arXiv:0911.3553]

\bibitem{un2}
L. Freidel and E.R. Livine,
{\it U(N) Coherent States for Loop Quantum Gravity},
J. Math. Phys. {\bf 52} (2011) 052502 [arXiv:1005.2090]

\bibitem{un3}
M. Dupuis and E.R. Livine,
{\it Revisiting the Simplicity Constraints and Coherent Intertwiners},
Class. Quant. Grav. {\bf 28} (2011) 085001 [arXiv:1006.5666]

\bibitem{un4}
M. Dupuis and E.R. Livine,
{\it Holomorphic Simplicity Constraints for 4d Spinfoam Models},
Class. Quant. Grav. {\bf 28} (2011) 215022 [arXiv:1104.3683]

\bibitem{un4_conf}
M. Dupuis and E.R. Livine,
{\it Holomorphic Simplicity Constraints for 4d Riemannian Spinfoam Models},
Proceedings of the Loops'11 conference 2011 (Madrid, Spain) [arXiv:1111.1125]

\bibitem{spinor}
E.F. Borja, L. Freidel, I. Garay and E.R. Livine,
{\it U(N) tools for Loop Quantum Gravity: The Return of the Spinor},
Class.Quant.Grav. {\bf 28} (2011) 055005 [arXiv:1010.5451]

\bibitem{spinor_johannes}
E.R. Livine and J. Tambornino,
{\it Spinor Representation for Loop Quantum Gravity},
arXiv:1105.3385

\bibitem{recursion3d}
V. Bonzom and L. Freidel,
{\it The Hamiltonian constraint in 3d Riemannian loop quantum gravity},
Class. Quant. Grav. {\bf 28} (2011) 195006 [arXiv:1101.3524]

\bibitem{recursion4d}
V. Bonzom,
{\it Spin foam models and the Wheeler-DeWitt equation for the quantum 4-simplex},
Phys. Rev. D {\bf 84} (2011) 024009 [arXiv:1101.1615]

\bibitem{2vertex}
E.F. Borja, J. Diaz-Polo, I. Garay and E.R. Livine,
{\it Dynamics for a 2-vertex Quantum Gravity Model},
Class.Quant.Grav.27 (2010) 235010 [arXiv:1006.2451]

\bibitem{ls}
E.R. Livine and S. Speziale,
{\it A new spinfoam vertex for quantum gravity},
Phys.Rev.D {\bf 76} (2007) 084028 [arXiv:0705.0674]

\bibitem{holquantumtet}
  L.~Freidel, K.~Krasnov and E.~R.~Livine,
  {\it Holomorphic Factorization for a Quantum Tetrahedron},
  Commun.\ Math.\ Phys.\  {\bf 297}, 45 (2010)
  [arXiv:0905.3627 [hep-th]].

\bibitem{sfcosmo}
E.R. Livine and M. Mart\'in-Benito,
{\it Classical Setting and Effective Dynamics for Spinfoam Cosmology},
arXiv:1111.2867


\bibitem{eprl}
J. Engle, E.R. Livine, R. Pereira and C. Rovelli,
{\it LQG vertex with finite Immirzi parameter},
Nucl. Phys. B {\bf 799} (2008) 136-149 [arXiv:0711.0146]

\bibitem{fk}
L. Freidel and K. Krasnov,
{\it A New Spin Foam Model for 4d Gravity},
Class. Quant. Grav. {\bf 25} (2008) 125018 [arXiv:0708.1595]

\bibitem{ls2}
E.R. Livine and S. Speziale,
{\it Consistently Solving the Simplicity Constraints for Spinfoam Quantum Gravity},
Europhys. Lett. {\bf 81} (2008) 50004 [arXiv:0708.1915].


\end{thebibliography}
\end{document}